\documentclass[twocolumn,showkeys,showpacs,preprintnumbers,prd,superscriptaddress,nofootinbib]{revtex4-1}
\usepackage{natbib}
\usepackage{graphicx}
\usepackage{epsf}
\usepackage{bm}
\usepackage{amsmath}
\usepackage{amsfonts}
\usepackage{amssymb}
\usepackage{epstopdf}
\usepackage{hyperref}
\usepackage{color}
\usepackage{verbatim}
\usepackage{multirow}
\usepackage{bm}
\usepackage{xcolor}
\usepackage[normalem]{ulem}
\usepackage{float}
\usepackage{placeins}

\begin{document}

%f+sigma8
%Probing Modified Gravity models with current cosmological data \\
\title{
\textcolor{black}{Modified Gravity from growth data: goodness-of-fit and gravitational coupling}\\
}

\author{Manoel V. S. Filho}
\email{manoelfilho@on.br}
\affiliation{Observatório Nacional, Rua General José Cristino 77, 
São Cristóvão, 20921-400 Rio de Janeiro, RJ, Brazil}

\author{Fernanda Oliveira}
\email{fernandaoliveira@on.br}
\affiliation{Observatório Nacional, Rua General José Cristino 77, 
São Cristóvão, 20921-400 Rio de Janeiro, RJ, Brazil}

\author{Wiliam S. Hipólito-Ricaldi}
\email{wiliam.ricaldi@ufes.br}
\affiliation{Grupo de Física Teórica e Computacional, CEUNES, 
Universidade Federal do Espírito Santo, Rodovia BR 101 Norte, 
km. 60, São Mateus, 29932-540, ES, Brazil} 
\affiliation{Núcleo Cosmo-UFES, CCE, Universidade Federal do Espírito Santo, Av. Fernando Ferrari, 540, CEP 29.075-910, Vitória, ES, Brazil}

\author{Felipe Avila}
\email{fsavila2@gmail.com}
\affiliation{Observatório Nacional, Rua General José Cristino 77, 
São Cristóvão, 20921-400 Rio de Janeiro, RJ, Brazil}

\author{Armando Bernui}
\email{bernui@on.br}
\affiliation{Observatório Nacional, Rua General José Cristino 77, 
São Cristóvão, 20921-400 Rio de Janeiro, RJ, Brazil}

\begin{abstract}
While background expansion data alone cannot discriminate among cosmological models, the growth of large-scale  structures directly probes the underlying theory of gravity, 
offering a path to distinguish General Relativity from its modifications. In this work, we constrain three representative $F(R)$ modified gravity models (Starobinsky, Hu-Sawicki, and 
$R^2$-corrected Appleby--Battye) using measurements of the growth rate $f(z)$ and the matter 
fluctuation amplitude $\sigma_8(z)$. Our analyses combine MCMC parameter estimation, Gaussian Process reconstructions, and goodness-of-fit statistics including Akaike information criterion (AIC) and Bayesian information criterion (BIC). All investigated models provide  statistically comparable fits to the current growth data, with information criteria differences too small to establish a preference for any particular scenario. To overcome this degeneracy, we reconstruct the effective gravitational 
coupling $\mu(z)$ from the MCMC posterior samples, providing a physically motivated diagnostic that complements standard goodness-of-fit criteria. The Starobinsky and Hu-Sawicki 
models predict moderate departures from General Relativity, remaining compatible with constraints on both $\mu(z)$ and $S_8$. In contrast, the $R^2$-AB model predicts 
$\mu_0 \sim 3.5$ and $S_8 = 0.638$, a $3.2\sigma$ tension with Planck 2018, rendering it physically disfavored despite its competitive statistical performance. 
\end{abstract}


\maketitle

%%------------------------------------------------
\section{Introduction}\label{introduction}

%We are interested in testing diverse cosmological scenarios using data from the perturbed universe.
%
%is questioning whether the $\Lambda$CDM model is the best model to describe the observed universe. Nothing is more timely than carrying out exhaustive analyses to confirm, or not, the performance of competing models in adequately describing the observed universe. And, concomitantly, new data from the clumpy universe are currently available. 
%can provide insight in the search of the final standard cosmological model. 

Accurate analyses of the Dark Energy Spectroscopic Instrument (DESI) baryon acoustic oscillation (BAO) measurements, together with cosmic microwave background (CMB) data and Type Ia supernova (SNIa) catalogs, have shown that the concordance cosmological model, flat-$\Lambda$CDM, is %increasingly 
less favored than  models with evolving dark energy, such as the $\omega_0\omega_a$CDM~\citep{DESI2025, DESI2025b}. 
These recent results have motivated extensive analyses of alternative cosmological scenarios, including extensions  of 
$\Lambda$CDM and modified gravity (MG) theories,  confronting them with current observational data~\citep{Oliveira2025a, Ribeiro23, Bessa2021, Nunes2017, Kumar2025, Luongo2024, Afroz2025, Euclid2021,DES2024}.

In this work, we perform a comparative analysis of representative $F(R)$ modified gravity models by studying the  evolution of matter perturbations over a wide redshift range. 
In particular, we focus on two cosmological observables: the growth rate of cosmic structures, $f(z)$, and the amplitude of matter density fluctuations on $8 \,h^{-1} 
\text{Mpc}$ scale, $\sigma_8(z)$. 
These observables provide complementary information on the evolution of matter clustering and are directly sensitive to the gravitational interaction responsible for structure formation. 
In fact, observables of the clumpy universe are particularly relevant for models in which the mechanism driving the clustering of matter structures is not based on General Relativity (GR)~\cite{Basilakos2013, Nesseris2017, Bessa2021, Bertini2020, Hipolito2025, Ribeiro23, DeFelice2010, Ishak2024}. This enables us to assess possible deviations from the flat-$\Lambda$CDM model and  to quantify how different  scenarios reproduce the observed growth of cosmic structures~\cite{Ribeiro23, Bessa2021, Nunes2017, Oliveira2025b, Artis2024, Kou2024, Chen2019, Basilakos2013, Basilakos2017, Xie2025, Luongo2024, Oliveira2026}. 
Among these scenarios, $F(R)$ theories provide a well-motivated framework in which departures from GR arise through an effective gravitational coupling that modifies the growth of matter perturbations.
%both data providing complementary information of the perturbed universe, allowing us to test deviations from th%e $\Lambda$CDM model. 
%In fact, the analyses we perform using the uncorrelated datasets $\{ f(z) \}$ and $\{ \sigma_8(z) \}$ are novel in the study of models with modified gravitational 
%interaction, that is, 
%models where the mechanism that drives the clustering of matter structures is not based on 
%General Relativity 
%(GR). 
%will help to break the degeneracy present in the commonly used  $[f\sigma_8](z)$ data, enabling 
%
%\textcolor{red}{[refs.]} 
%for a more detailed assessment of the mechanisms driving the formation of cosmic structures ~\cite{Franco2025, Avila21} 
%(\textcolor{blue}{Isto é verdade? foi mostrado?}), 

%In this context, most previous analyses have either constrained the parameters of specific MG theories or adopted model-independent phenomenological parameterizations of the effective gravitational coupling. In this work, we combine these two complementary strategies within a unified framework, allowing a direct comparison between theoretically motivated $F(R)$ models and phenomenological constraints on deviations from GR.

We firstly compute the theoretical evolution of $f(z)$ and $\sigma_8(z)$ for each $F(R)$ model and constrain their parameters through a Markov Chain Monte Carlo (MCMC) analysis using current growth measurements. This provides
the main statistical inference of the work and allows the different models to be consistently compared through their posterior constraints, goodness-of-fit, and information criteria. 
%From the perspective of linear perturbation theory, cosmological observables related to the growth of structures depend both on the background expansion and on the underlying theory of gravity \cite{Linder2005}. 
%
Specifically, we study the Starobinsky~\cite{Starobinsky2007}, Hu-Sawicki~\cite{Hu2007}, 
and $R^2$-corrected Appleby--Battye ($R^2$-AB)~\cite{Appleby2007} 
$F(R)$ models, which predict distinct evolutions for $f(z)$ and $\sigma_8(z)$. 
At the linear perturbation level, deviations from GR are encoded in the effective gravitational coupling $\mu(z,k)$, where $\mu = \text{const.} = 1$ corresponds to a description based on GR theory, and departures from unity represent modifications on cosmological scales through the modified Poisson equation. 
These deviations can be directly tested against current observational data to assess whether tensions reported in the literature may signal new physics~\cite{macaulay2013, Planck2018, Perivolaropoulos2022, DiValentino2025, Nunes2021, Avila21, Avila22a, Adil2024, Franco25b}.  
Recent analyses, for instance using the first-year DESI clustering data along with Planck CMB, CMB lensing  from Planck and ACT, Big Bang Nucleosynthesis constraints, DES Y3  weak lensing and clustering, and DES Y5 supernovae, have placed bounds on late-time deviations from GR ~\cite{Ishak2024}.

Beyond the statistical comparison provided by the MCMC analysis, we reconstruct the effective gravitational coupling $\mu(z,k)$ from the posterior samples of each model. This provides a direct diagnostic of the modification of the gravitational interaction predicted by the underlying theory, complementing the information obtained from standard goodness-of-fit statistics. 
We further compare the derived $S_8$ values with
the \textit{Planck} 2018 determination, providing an additional test of the consistency of the inferred growth of structures. 
As a complementary, model-independent benchmark, Gaussian Process (GP) reconstructions of $f(z)$ and $\sigma_8(z)$ are also considered, serving as a reference for the evolution of observables describing the growth of cosmic structures and not as an additional source of parameter constraints.

The main objective is therefore not simply to identify the model with the lowest $\chi^2$, but to determine whether statistically competitive $F(R)$ scenarios also
predict a gravitational interaction and clustering amplitude compatible with current
observational constraints. This combined analysis allows models with similar
statistical performance to be physically discriminated through their predicted
$\mu(z,k)$ and $S_8$.

This work is organized as follows. In Section~\ref{basics}, we present the basic equations governing matter perturbations, including $f(z)$ and $\sigma_8(z)$. Section~\ref{data} describes the cosmological observables and the data compilation used in our analyses. Section~\ref{methodology} details the methodology, including the MCMC technique used to constrain the model parameters. Section~\ref{models} introduces the $F(R)$  MG models considered in this work. Finally, Section~\ref{results} presents and discusses the results of our comparative analysis, combining statistical model comparison (AIC/BIC), the reconstructed effective gravitational coupling $\mu$, and a comparative analysis of $S_8$ against the Planck 2018 reference. We leave to the Appendices the details of the GP methodology (Appendix~A), the functional forms and viability conditions of the alternative cosmological models (Appendix~B), and supplementary material on the GR-based extensions and the $S_8$ tension calculation (Appendix~C).

%--------------------------------------------
%\section{Cosmological observables and data}
\section{Cosmological observables of matter perturbations}
\label{basics}

The evolution of matter clustering and the growth of cosmic structures 
is described using linear perturbation theory, studying the time evolution of 
the matter density contrast 
$\delta_m(t, \textbf{r})$~\cite{Coles1996, Avila21, Marques20, Franco25b}, 
a quantity defined at position $\textbf{r}$, at cosmic time $t$. 
%(sometimes the scale factor $a(t)$ is used instead of the time variable $t$). 
Working in Fourier space, where $k$ denotes the comoving wavenumber, for MG theories at sub-horizon scales and in the quasi-static approximation (i.e., $k^2/a^2 \gg H^2$), the evolution of the matter density contrast is governed by the second-order differential equation~\citep{Ribeiro23} 
\begin{equation}\label{eq:edo}
\ddot{\delta}_m(t) + 2 H(t) \dot{\delta}_m(t) - 4 \pi G_{\text{eff}}(t,k)\,\bar{\rho}_m(t)\, \delta_m(t) = 0 \,,
\end{equation}
where the dots denote derivatives with respect to time $t$, 
and $H(t)\equiv\dot{a}(t)/a(t)$ is the Hubble parameter. 
The function $G_{\text{eff}}(t,k) \equiv G_{N}\,\mu(t,k)$, where $G_{N}$ is the Newtonian gravitational constant, defines the effective gravitational coupling, $\mu(t,k)$, a function that  encodes possible deviations from GR. 
In the frame of GR theory, one has $\mu(t,k) = 1$, while in MG theories $\mu(t,k)$ is scale and time dependent. 
%?
Because the scale factor $a$ is related to the redshift $z$, $a(t) = 1/(1+z)$, one can also write $\mu(t,k) = \mu(a,k) = \mu(z,k)$. 
%?

In fact, in the context of $F(R)$ gravity, the presence of a scalar degree of freedom alters the effective gravitational interaction, $G_{\rm eff}$. 
From the analysis of  sub-horizon scalar perturbations in  quasi-static regime, $G_{\rm eff}$ can be expressed as \cite{Tsujikawa2008, Pogosian2008, DeFelice2010}
\begin{equation}\label{mu}
\mu(a,k) = \frac{G_{\rm eff}(a,k)}{G_N} = \frac{1}{F'(R)} \,\frac{1 + 4\,p \,(k^2 / a^2 R)}{1 + 3\,p \,(k^2 / a^2 R)} \,,
\end{equation}  
where 
\begin{equation}
p \equiv \frac{R\,F''(R)}{F'(R)} \,,
\end{equation}  
with $F'(R) \equiv dF(R)/dR$. In the GR limit, $F(R) \to R$ implies $F'(R) \to 1$ and
$F''(R) \to 0$, so that $p \to 0$ and equation~(\ref{mu}) correctly 
reduces to $\mu = \text{const.} = 1$, recovering the standard GR behavior. 
This formulation captures the scale-dependent modifications of gravity induced by the extra scalar degree of freedom, i.e., the scalaron, present in the $F(R)$ gravity. 
To ensure compatibility with the linear regime probed by current growth measurements, we adopt a representative scale of $k = \bar{k} = 0.125\, h\,\text{Mpc}^{-1}$, in agreement with previous analyses~\cite{Ribeiro23, Kumar2025, Basilakos2013, Tsujikawa2008}.  %From now on, we write $\mu(a)$ or $\mu(z)$ instead of $\mu(a,\bar{k})$ or $\mu(z,\bar{k})$, respectively. 

In the regime of linear perturbations, the density contrast is a function of time only. Consequently, we can define the growth rate of cosmic structures
\begin{equation}\label{defgrowthrate}
f(a) \equiv \frac{d \hspace{0.01cm} \ln \delta(a)}{d \hspace{0.01cm} \ln a} \,.
\end{equation}

%An useful parametrization for the growth rate of cosmic structures $f(z)$ is given by~\cite{Wang1998, Amendola2004, Linder2005}
%\begin{equation}
%\label{eqLinder}
%f(a) = \Omega_m ^{\,\gamma} (a),
%\end{equation}
%where $\gamma$ is the growth index and $\Omega_m(a)$ is the matter-energy density function. 
%In GR, the value of the growth index is a constant 
%the $\Lambda$CDM model, 
%$\gamma = 6/11 \simeq 0.55$. 
%In MG and alternative cosmological theories, this parameter can assume different 
%values~\cite{Basilakos2012, Linder2007}.

%It is important to note that this parametrization is obtained considering an euclidean universe. 
%For spacetimes with $\Omega_k \ne 0$, one has to use~\cite{Gong2009}
%\begin{equation} 
%\label{fcurved}
 %   f(a) = \Omega_m ^{\,\gamma} (a) + \left(\gamma - \frac{4}{7} \right)\Omega_k(a),
%\end{equation}
%where $\Omega_k(a)$ is the curvature-energy density function.

The other cosmic observable considered for our analyses is the matter fluctuations amplitude at the scale of $8$ Mpc$/h$, $\sigma_8(a)$, equivalently $\sigma_8(z)$, is given by~\cite{Ribeiro23, Nesseris2017}
\begin{equation}
\label{sig8}
\sigma_8(a) \equiv \sigma_{8,0} \left[\frac{\delta_m(a)}{\bar{\delta}_m(1)} \right] \,,    
\end{equation}
where $\sigma_{8,0} \equiv \sigma_8(a=1)$, equivalently $\sigma_{8,0} = \sigma_8(z=0)$), 
and $\bar{\delta}_m(a=1)$ is the normalization factor obtained by integrating the growth equation, that is, equation~(\ref{eq:edo}). 

\section{Data}\label{data}

We consider 11 measurements of  the growth rate of cosmic structures, denoted as $f(z)$, compiled in~\cite{Avila22b} and presented in Table \ref{tab:table0}; and 14 measurements 
of $\sigma_8(z)$ compiled by~\cite{Piccirilli2024}, plus 1 recent measurement at low-redshift by~\cite{Franco2025}, detailed in Table \ref{tab:table1}. 
The $f(z)$ compilation follows the selection criteria of ~\citep{Avila22b}, retaining only direct measurements of $f$ (rather than $f\sigma_8$ value converted to $f$ through a fiducial cosmology) and using uncorrelated redshift bins for a given cosmological tracer, which minimizes internal covariance within this dataset. 
The $\sigma_8(z)$ measurements, in turn, are obtained independently through different methodologies applied to diverse cosmic tracers, 
as for instance CMB lensing cross-correlation analyses of quasar catalogs, galaxy clustering, or cosmic shear measurements~\citep{Piccirilli2024, Miyatake2022, Abbott2023, ACT2023, Abbott-Kilo-DegreeSurvey, Belsunce2025, Franco2025}. 
Given the lack of correlation between the $\{f(z_i)\}$ and $\{\sigma_8(z_i)\}$ datasets, in terms of survey systematics, tracers, and estimators, we adopt a diagonal covariance matrix in the joint likelihood analysis, following similar treatments in the literature \citep{Nunes2021, Garcia2021, DiValentino2021}.

%2103.15862 H. Miyatake et al.
%2105.12108 C. Garcia-Garcia et al.
%2207.05766 T.M.C. Abbott et al.   
%2309.05659  ACT collaboration
%2305.17173 Kilo-Degree Survey, Dark Energy Survey collaboration, DES Y3 + KiDS-1000:
\begin{table}[h]
\caption{Dataset of 11 measurements of $f(z)$, from~\cite{Avila22b}.}
\centering
\begin{tabular}{ccc@{\hspace{1.5em}}ccc}
\hline\hline
$z$ & $f(z)$ & error & $z$ & $f(z)$ & error \\
\hline
0.013 & 0.56 & 0.07 & 0.41 & 0.70 & 0.07 \\
0.150  & 0.49 & 0.14 & 0.55 & 0.75 & 0.18 \\
0.180  & 0.49 & 0.12 & 0.60 & 0.73 & 0.07 \\
0.220  & 0.60 & 0.10 & 0.77 & 0.91 & 0.36 \\
0.350  & 0.70 & 0.18 & 1.40 & 0.90 & 0.24 \\
0.380  & 0.66 & 0.09 &      &      &      \\
\hline\hline
\end{tabular}
\label{tab:table0}
\end{table}

%\begin{table}[h]
%\caption{Dataset of 11 measurements of 
%$f(z)$, from~\cite{Avila22b}.}
 %   \begin{tabular}{ c | c | c}
%\,\,$z$\,\,  & \,\,$f(z)$\,\, & \,\,error    \\
%\hline
%$0.013$ &  $0.56$  &  $0.07$  \\   
%$0.15$  &  $0.49$  &  $0.14$  \\     
%$0.18$  &  $0.49$  &  $0.12$  \\  
%$0.22$  &  $0.60$  &  $0.10$  \\  
%$0.35$  &  $0.70$  &  $0.18$  \\   
%$0.38$  &  $0.66$  &  $0.09$  \\  
%$0.41$  &  $0.70$  &  $0.07$  \\   
%$0.55$  &  $0.75$  &  $0.18$  \\     
%$0.60$  &  $0.73$  &  $0.07$  \\  
%$0.77$  &  $0.91$  &  $0.36$  \\  
%$1.40$  &  $0.90$  &  $0.24$  \\   
%\end{tabular}
%\label{tab:table0}
%\end{table}

\begin{table}[h]
\caption{Dataset of $15$ measurements of $\sigma_8(z)$ from~\cite{Piccirilli2024, Franco2025}.}
\centering
\begin{tabular}{ccc@{\hspace{1.5em}}ccc}
\hline\hline
$z$ & $\sigma_{8}(z)$ & error & $z$ & $\sigma_{8}(z)$ & error \\
\hline
0.013 & 0.78 & 0.04 & 0.83 & 0.58 & 0.04 \\
0.240  & 0.67 & 0.04 & 0.92 & 0.44 & 0.06 \\
0.470  & 0.58 & 0.04 & 1.10 & 0.48 & 0.01 \\
0.530  & 0.59 & 0.03 & 1.50 & 0.46 & 0.05 \\
0.600  & 0.59 & 0.02 & 1.59 & 0.39 & 0.06 \\
0.630  & 0.53 & 0.04 & 2.72 & 0.22 & 0.06 \\
0.690  & 0.66 & 0.10 & 3.80 & 0.12 & 0.06 \\
0.800  & 0.47 & 0.04 &      &      &      \\
\hline\hline
\end{tabular}
\label{tab:table1}
\end{table}

%\begin{table} [h]
%\caption{Dataset of $15$ measurements of 
%$\sigma_8(z)$ from~\cite{Piccirilli2024, Franco2025}.}
%\begin{tabular}{l | c | c}
%$z$  & \,\,$\sigma_{8}(z)$\,\, & error \\ \hline
%$0.013$ & 0.78 & 0.04 \\ \hline
%0.24 & 0.67            & 0.04 \\ \hline
%0.47 & 0.58            & 0.04 \\ \hline
%0.53 & 0.59            & 0.03 \\ \hline
%0.60 & 0.59            & 0.02 \\ \hline
%0.63 & 0.53            & 0.04 \\ \hline
%0.69 & 0.66            & 0.10 \\ \hline
%0.80 & 0.47            & 0.04 \\ \hline
%0.83 & 0.58            & 0.04 \\ \hline
%0.92 & 0.44            & 0.06 \\ \hline
%1.10 & 0.48            & 0.01 \\ \hline
%1.50 & 0.46            & 0.05 \\ \hline
%1.59 & 0.39            & 0.06 \\ \hline
%2.72 & 0.22            & 0.06 \\ \hline
%3.80 & 0.12            & 0.06 \\ 
%\end{tabular}
%\label{tab:table1}
%\end{table}

%In modified gravity models, $Q(t)$ can be written as~\cite{Ribeiro23}
%\begin{equation}
%Q(t) = \frac{1}{F'(R)}\,
%\frac{1 + 4p\!\left(\frac{k^2}{a^2 R}\right)}
%{1 + 3p\!\left(\frac{k^2}{a^2 R}\right)} \,,
%\label{eq:Qt}
%\end{equation}
%\textcolor{red}{where $R$ is the spacetime curvature, $F'(R)$ is ..., $p$ is..., $k$ is ..., $R$ is ...}
%Solutions of equation~(\ref{eq:edo}) depend on the specific theory of the MG assumed. 
In addition, we consider the cosmological constraints on deviations of GR 
reported in~\citep{Ishak2024} analyzing the first-year of clustering observations from DESI in combination with other available datasets including the CMB data from Planck with CMB-lensing from Planck and ACT collaborations, BBN constraints on the physical baryon density, the galaxy weak lensing and clustering from DESY3, and supernova data from DESY5 \cite{Ishak2024}. 
Using the functional parameterization 
\begin{equation}\label{eq:desi_mu}
\mu_{\rm DESI}(z) = 1 + \Delta_0 \, \frac{\Omega_{\rm DE}(z)}{\Omega_\Lambda} \,,
\end{equation}
where $\Omega_{\rm DE}(a) \equiv (8\pi G\,\rho_{\rm DE}(z))/3H^2(z)$
is the fractional dark energy density at redshift $z$, and 
the parameter $\Delta_0$ is equal to zero in GR. 
%$\Delta_0 = 0$ corresponds to GR. 
The combination of datasets DESI(FS+BAO)+CMB+DES-Y3+DES-Y5-SN yields 
\begin{equation}
\Delta_0 = 0.05 \pm 0.22 \,,
\label{eq:desi_mu0_value}
\end{equation}
in the flat-$\Lambda$CDM background~\cite{Ishak2024}. 
Note that $\Delta_0=0$ implies that $\mu_{\rm DESI} = \text{const.} = 1$, 
which corresponds to GR. 
%Because $\Omega_{\rm DE}(z=0)/\Omega_\Lambda = 1$ by construction, the 
At $z=0$, the DESI parametrization, given in equation~(\ref{eq:desi_mu}), 
can be related with our $\mu$, 
\begin{equation}
\mu_{\rm DESI}(z=0) - 1 \equiv \mu_0 - 1 = 0.05 \pm 0.22 
\label{eq:desi_deltamu_z0}
\end{equation}
where we have defined $\mu_0 \equiv \mu(z=0)$. 
Note that this relationship does not hold at other redshifts, since 
$\mu_{\rm DESI}(z)-1 = \Delta_0\, \Omega_{\rm DE}(z)/\Omega_{\Lambda}$ is itself a function of $z$, not a constant offset.

The choice for adopting the fixed scale $k = \bar{k} = 0.125\, h\,\text{Mpc}^{-1}$ is reasonable. 
To show this, we evaluate the difference $\delta\mu \equiv \mu(z=0,k_{\rm max})  - \mu(z=0,k_{\rm min})$ across the DESI wavenumber range $0.02 \leq k \leq 0.20\,h\,{\rm Mpc}^{-1}$~\cite{Ishak2024}, using representative parameter values consistent with local gravity constraints together with Planck~2018 cosmological parameters~\cite{Planck2018}, we found: 
for the Hu-Sawicki model ($c_2 \in [10,100]$, $n=1,2$), 
$\delta\mu < 1.4 \times 10^{-5}$, confirming negligible scale dependence; 
for the  Starobinsky model ($\lambda_S \in [0.5,2.0]$, $n=1,2$), 
$\delta\mu < 3.1 \times 10^{-3}$, less than $2.1\%$ of $(\mu_0-1)$; 
and for the R$^2$-AB model ($b \in [1.6,4.0]$), 
$\delta\mu < 3.7 \times 10^{-3}$, a relative  variation below $2.2\%$ of $(\mu_0-1)$. 
In all cases, the scale dependence within the current observational range is subdominant in relation to the measurement uncertainties, justifying the 
adoption of a fixed representative scale $\bar{k}$.

%%%%%%%%%%%%%%%%%%%%%%%%%%%%%%%%%%%%%%%%%%
\section{Methodology} \label{methodology}
%%%%%%%%%%%%%%%%%%%%%%%%%%%%%%%%%%%%%%%%%%
Our  analyses consist of three main steps. 
For each MG model introduced in Section~\ref{models}, we first compute the theoretical predictions for the growth observables $f(z)$ and $\sigma_8(z)$ by numerically solving equation~(\ref{eq:edo}). 
In the second step, we perform a MCMC sampling \cite{Gilks1995, Gelman2013} using these observables, adopting flat priors on all free parameters: $H_0 \in [60, 80]$, $\Omega_{m0} \in [0.1, 0.5]$, and $\sigma_8 \in [0.5, 1.2]$, while the additional parameters are constrained within model-dependent intervals: %$\Omega_{k0} \in [-0.2, 0.5]$ for the open curvature model, $w_0 \in [-2, -0.3]$ and $w_a \in [-2, 2]$ for the CPL parametrization, 
$c_2 \in [10, 200]$ for the Hu-Sawicki models, $\lambda_s \in [0.1, 2.0]$ for the Starobinsky models, and $b \in [0.5, 5.0]$ for the $R^2$-AB model, whose definitions are given in Appendix \ref{altenative-models}. 
The statistical inference is based on the likelihood function
\begin{equation}
\mathcal{L} \propto \exp\!\left(-\frac{\chi^{2}}{2}\right),
\end{equation}
where the chi-square statistic is 
\begin{equation}
    \chi^{2} = \sum_{ij} \Delta E_{i}\, C^{-1}_{ij}\,\Delta E_{j},
\end{equation}
with residuals
\begin{equation}
    \Delta E_{i} = E_{i}(\vartheta|\alpha) - D_{i}.
\end{equation}
with $\vartheta$ denoting the vector of free parameters varied in the MCMC, $E_{i}(\vartheta|\alpha)$ denotes the theoretical prediction for the $i$-th observable, $D_{i}$ is the corresponding observational measurement, and $C_{ij}$ the covariance matrix. The MCMC sampling is performed using the affine-invariant ensemble sampler implemented in the \texttt{emcee} package \cite{ForemanMackey2013}, which employs the Goodman--Weare stretch move algorithm. For each model, the chains are generated with 50-60 walkers evolving over 5000--6000 steps, where the first 1000 steps are discarded as burn-in to allow the sampler to reach the stationary regime. Convergence is assessed through visual inspection of the trace plots together with an analysis of the integrated autocorrelation time $\tau$, which is verified to remain well below the total chain length, ensuring that the retained samples are effectively independent. 
To further quantify the relative performance of the models, we employed the AIC and BIC, which assess the trade-off between the quality of fit and model complexity. 
The AIC is defined as 
\begin{equation} 
\mathrm{AIC} \equiv \chi^2_{\min} + 2K \,\text{,}
\end{equation}
and the BIC as 
\begin{equation}
\mathrm{BIC} \equiv \chi^2_{\min} + K \ln N \,\text{,}
\end{equation}
where $K$ represents the number of free parameters and \(N\) the number of observational data points.

In the third step, we reconstruct the effective MG function $\mu(z,\bar{k})$ 
{\it a posteriori} using the MCMC posterior samples of the model parameters. This procedure ensures that any inferred deviation from GR arises self-consistently from the underlying gravitational model. Since the model parameters are generally correlated within the posterior distribution, the uncertainty on $\mu(z,\bar{k})$ cannot be obtained by propagating the individual parameter uncertainties independently. 
Instead, $\mu(z,\bar{k})$ is computed for every posterior sample of the MCMC chains, fully preserving the parameter correlations encoded in the posterior distribution. The mean evolution of $\mu(z,\bar{k})$, together with its 1$\sigma$ credible regions at each redshift, is obtained directly from the resulting ensemble. We then compare the reconstructed $\mu(z,\bar{k}) - 1$ for each model. 
This comparison allows us to assess whether the deviation from GR predicted by each MG model remains consistent with current growth-rate and DESI constraints simultaneously.

%
%The reconstructed $\mu(a)$, or $\mu(z)$, is then compared with phenomenological constraints obtained from model-independent analyses, allowing us to assess the consistency of the theoretical model with observational bounds on deviations from GR. This procedure is repeated independently for each model considered.

%Our analysis is performed considering one specific MG model at a time. 
%Model parameters are constrained through a Markov Chain Monte Carlo (MCMC) analysis using %large-scale structure observables. 
%From the resulting posterior distributions, we reconstruct the effective MG function $\mu(a)$, which encodes 
%--for fixed scale $k = 0.125\, h \text{Mpc}^{-1}$-- the time-dependence of the modified gravitational interaction. 

%Finally, from the best-fit parameters obtained in the MCMC, we reconstruct the function $\mu(a)$ characterizing deviations from GR, and then compare its present time values with the empirical constraints reported by DESI and complementary datasets, namely $\mu_{0}=0.05 \pm 0.22$ for time-only parameterizations, as well as the binned constraints $\mu_{1}$ and $\mu_{2}$ in redshift space. 
%This enables a direct and quantitative assessment of whether the models studied remain compatible with current measurements of large-scale structure growth and its consistency with the predictions of GR theory. 

As a complementary and model-independent consistency check, we also apply GP regression directly to the $f(z)$ and $\sigma_8(z)$ datasets (See appendix \ref{appendixA} for details). 
This non-parametric reconstruction provides an independent benchmark for the cosmic growth observables, avoiding any assumption regarding the underlying cosmological model, gravitational theory, or expansion history. 
Consequently, it enables an unbiased comparison between the observational data and the predictions of all cosmological models considered in this work. As such, the GP regression is configured to yield both 1$\sigma$ and 2$\sigma$ confidence bands for each structure growth observable, providing a statistically robust, theory-agnostic reference against which the goodness of fit of every parametric model can be evaluated.

\section{Studied models} \label{models}
We consider three representative $F(R)$ MG models: Starobinsky, Hu-Sawicki, and $R^2$-corrected
Appleby--Battye ($R^2$-AB). Each model modifies the gravitational sector by introducing, in addition to the standard cosmological parameters $(H_0,\Omega_{m0},\sigma_8)$, a small number of parameters governing the functional form of $F(R)$. The complete functional forms and free parameters are summarized  in Table~\ref{tab:mg_summary}. Stability conditions, and de--Sitter viability requirements are discussed in Appendix~\ref{modelsfr}.

For the Starobinsky model, we consider the cases $n=1$ and $n=2$, leaving $\lambda_S$ as the only additional free parameter. The case $n=1$, differently from the case $n=2$, is known to have difficulty in passing solar system tests and reproducing the matter density power spectrum, nevertheless, it is still studied in the literature as a prototypical example of the theory~\citep{Starobinsky2007, DeFelice2010, Motohashi2009, Bessa2021}. For the Hu-Sawicki model, we likewise consider $n=1$ and $n=2$, with $c_2$ as the additional free parameter. 
These choices preserve the chameleon screening mechanism and are consistent with local gravity constraints~\citep{Hu2007, DeFelice2010, Tsujikawa2008}. 
For the $R^2$-AB model, the parameter $b$ controls the departure from the GR regime. Requiring $b\geq 1.6$ allows the model to reproduce the recent cosmic acceleration \citep{Appleby2007, Appleby2009, Ribeiro23}. 
The parameter $M$ sets the scalaron mass scale and is fixed by the amplitude of the primordial power spectrum, with $M\simeq 1.2\times10^{-5}M_{\rm P}$ during inflation \citep{motohashi, 
Starobinsky1980, Planck2018Inflation}.

\begin{table*}[t]
\caption{Summary of the $F(R)$ modified gravity models considered 
in this work. The last column lists the additional free parameter 
beyond the standard flat-$\Lambda$CDM cosmological parameters 
($H_0$, $\Omega_{m0}$, $\sigma_8$). Throughout this work, the 
exponent $n$ is fixed ($n=1$ or $n=2$), so it is not treated as 
a free parameter.}
\centering
\begin{tabular}{ccc}
\hline\hline
Model & $F(R)$ & Parameter \\
\hline
Starobinsky &
$\displaystyle F(R)= R+\lambda_S R_S
\left[\left(1+\frac{R^2}{R_S^2}\right)^{-n}-1\right]$ &
$\lambda_S$ \\[0.5cm]
Hu-Sawicki &
$\displaystyle F(R)= R - m_0^2
\frac{c_1\left(R/m_0^2\right)^n}{c_2\left(R/m_0^2\right)^n+1}$ &
$c_2$ \\[0.5cm]
$R^2$-AB &
$\displaystyle F(R)= \frac{R}{2}+\frac{\epsilon_{\rm AB}}{2}
\ln\!\left[\frac{\cosh(R/\epsilon_{\rm AB}-b)}{\cosh b}\right]
+\frac{R^2}{6M^2}$ &
$b$ \\[0.3cm]
\hline\hline
\end{tabular}
\label{tab:mg_summary}
\end{table*}

%\begin{table*}[t]
%\centering
%\small
%\setlength{\tabcolsep}{6pt}

%\caption{Summary of the $F(R)$ modified gravity models considered in this work. The last column lists the additional free parameter beyond the standard flat-$\Lambda$CDM cosmological parameters ($H_0$, $\Omega_{m0}$, $\sigma_8$). Throughout this work, the exponent $n$ is fixed ($n=1$ or $n=2$), so it is not treated as a free parameter.}

%\label{tab:mg_summary}

%\begin{tabular*}{0.90\textwidth}{@{\extracolsep{\fill}}lcl}
%\hline
%\textbf{Model} & $\mathbf{F(R)}$ & \hspace{-0.7cm}\textbf{Parameter} \\
%\hline \\

%Starobinsky &
%$\displaystyle
%F(R)=
%R+\lambda_SR_S
%\left[
%\left(1+\frac{R^2}{R_S^2}\right)^{-n}-1
%\right]
%$
%&
%$\lambda_S$
%\\[0.35cm]

%Hu-Sawicki &
%\hspace{-1.2cm}$\displaystyle
%F(R)=
%R-
%m_0^2
%\frac{c_1\left(R/m_0^2\right)^n}
%{c_2\left(R/m_0^2\right)^n+1}
%$
%&
%$c_2$
%\\[0.35cm]
%
%$R^2$-AB &
%\hspace{1.cm}$\displaystyle
%F(R)=
%\frac{R}{2}
%+
%\frac{\epsilon_{\rm AB}}{2}
%\ln\!\left[
%\frac{\cosh\!\left(R/\epsilon_{\rm AB}-b\right)}
%{\cosh b}
%\right]
%+
%\frac{R^2}{6M^2}
%$
%&
%$b$ \\ \\
%\hline
%\end{tabular*}
%\end{table*}

%%---------------------------------------------------------
\section{Results and Discussions}\label{results}

In this section we perform statistical analyses to determine the performance 
of MG models to describe the cosmic evolution of the observables $f(z)$ and 
$\sigma_8(z)$, reconstructed via GP using current data. 
Using the results of these analyses we perform a model comparison, with respect to the flat-$\Lambda$CDM model. 
In addition, we constrain the effective gravitational coupling, 
$\Delta\mu(z) \equiv \mu(z) - 1$, for each MG model investigated. 
Finally, we evaluate $S_8$ for each model, and perform a comparative study.

%?
%We use current cosmological data to perform comparative statistical analyses between the concordance cosmological model, the flat-$\Lambda$CDM, and alternative models based on GR theory and also MG models. 
%In particular, we focus on the observational data from the growth rate of cosmic structures, $f(z)$, and from the amplitude of matter density fluctuations on $8$ Mpc$/h$ scale, $\sigma_8(z)$, both data providing complementary information of the perturbed universe, allowing us to test deviations from the concordance model. 
%Our analyses using these uncorrelated datasets are novel in the study of possible modifications of the gravitational interaction, i.e., the mechanism that drives the formation of cosmic structures. 
%?
\subsection{MCMC constraints and Gaussian Process comparisons}
\begin{figure}[H]
\centering
\includegraphics[width=0.5\textwidth]{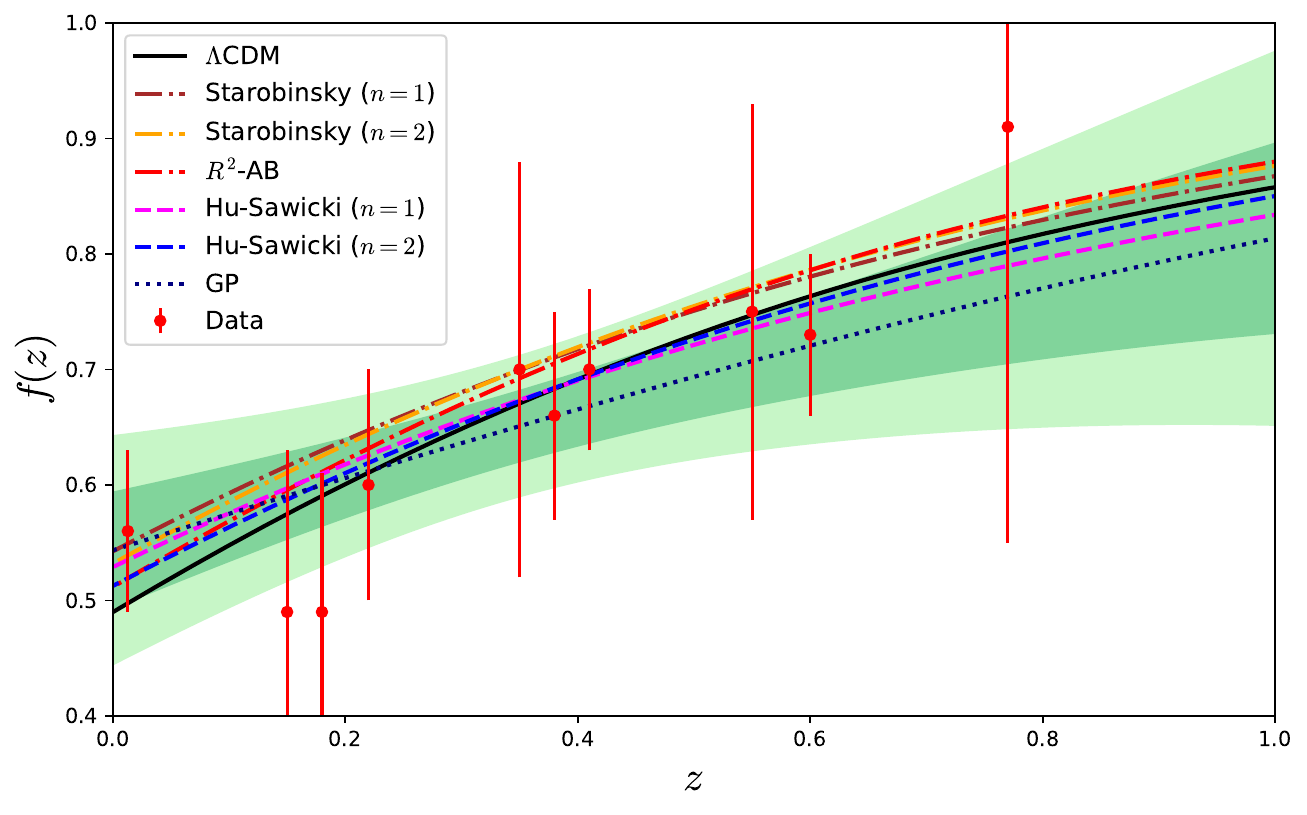}
\caption{Comparison among GP reconstruction of $f(z)$ and MG models.}
%Table~\ref{tab:table3}
\label{fig:fz_gm}
\end{figure}

\begin{figure}[h!]
\centering
\includegraphics[width=0.5\textwidth]{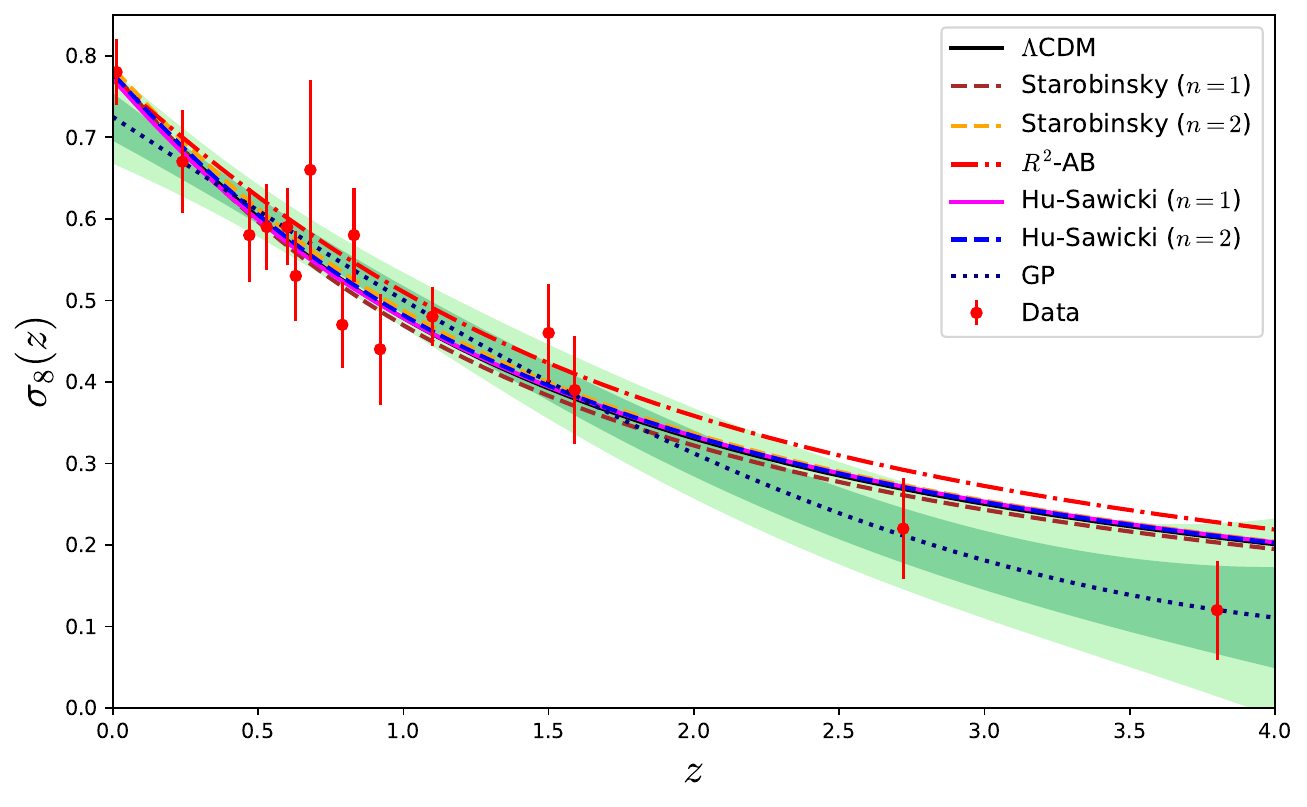}
\caption{Comparison among GP reconstruction of $\sigma_8(z)$ and MG models.}
\label{fig:sigma8_gm}
\end{figure}

Our first analysis combines parametric predictions, obtained through MCMC analyses, with model-independent GP reconstructions. This approach enables us to assess not only the statistical agreement between theory and observations, but also the physical consistency of the gravitational dynamics responsible for the growth of cosmic structures. The posterior constraints obtained from the MCMC results are summarized in Table~\ref{tab:MG_parameters}. 
These results provide the best-fit cosmological parameters used to generate the theoretical predictions displayed in Figures~\ref{fig:fz_gm} and \ref{fig:sigma8_gm}. 
The inferred values of $H_0$, $\Omega_{m0}$, and $\sigma_8$ remain broadly consistent among the different scenarios. 
The only exception is the $R^2$-AB model, which favors a lower 
matter density $\Omega_{m0} \simeq 0.205$, 
a feature that is connected to a low value of $S_8$ (as shown in Section~\ref{s8}). 
However, caution is needed here. 
This low value of $\Omega_{m0}$ may be connected with the use of the growth data used, 
since this same model tends to favor a higher value when other datasets are 
employed~\citep{Ribeiro23}.

Besides this feature, we observe that the present-day clustering amplitude is stable, 
with all models predicting $\sigma_8 \approx 0.77$. 
This indicates that the current growth measurements mainly constrain the redshift evolution of matter perturbations rather than the normalization of the matter power spectrum.

%Firstly, we have reconstructed the growth rate $f(z)$ and the clustering amplitude $\sigma_8(z)$ functions, obtained through GP applied to current observational data. 
%These non-parametric reconstructions allow us a comparison with predictions from both MG models and GR-based cosmologies, providing a comprehensive assessment of their ability to describe the growth rate of cosmic structures.

Figure~\ref{fig:fz_gm} compares the best-fit predictions of the MG models with the data and the GP reconstruction of the growth rate $f(z)$. 
In general, all the $F(R)$ models  reproduce the reconstructed growth history within the GP 2$\sigma$ confidence region, indicating a good agreement with the current observational data over the full redshift interval. 
%Among all scenarios, the Hu–Sawicki $n=1$ model follows the reconstructed evolution most closely across the observed $z$-range, while the $R^2$-AB model predicts a slightly faster growth at late times compared with the GP reconstruction. 
%In contrast, the standard flat-$\Lambda$CDM prediction tends to lie above the MG scenarios at higher redshifts, indicating a stronger growth of matter perturbations. 
%Although these 
The differences observed in this figure remain statistically modest given the current observational uncertainties. They just illustrate how the diversity of gravitational interactions in the investigated models can affect the evolution of the growth of cosmic structures, while remaining statistically compatible with present observations.

%Figure~\ref{fig:fz_gm} shows the GP reconstruction of $f(z)$ (dashed blue line) alongside MG models, including Starobinsky, $R^2$-AB, and Hu–Sawicki. 
%The reconstruction captures the observational trend at low and intermediate redshifts, with most MG models lying within the GP uncertainty band. 
%The Hu–Sawicki model ($n=1$) matches the reconstructed curve particularly well, indicating a slower but consistent growth of structures. 
%In contrast, the standard flat-$\Lambda$CDM prediction exceeds other curves at higher redshifts, reflecting a faster growth rate than that favored by the GP reconstruction or by some MG scenarios. 
%As observed, the MG models show consistency with the data, reconstructed using GP. 

Figure~\ref{fig:sigma8_gm} presents the corresponding comparison for the clustering amplitude $\sigma_8(z)$. A similar general behavior as in the $f(z)$ case is observed. The best-fit modified gravity models remain  consistent with both the observational measurements and the GP reconstruction, confirming that the current $\sigma_8(z)$ dataset does not provide strong discrimination among viable cosmological scenarios. Nevertheless, small differences appear over the redshift range covered by the data. 
The $\Lambda$CDM and  most of MG models tend to overestimate the clustering amplitude at redshifts $z \gtrsim 1.5$. 
The only MG model with a different behavior is the $R^2$-AB model which exhibits a larger amplitude across the entire $z$ range. 
This behavior is consistent with the enhanced effective gravitational coupling predicted by the $R^2$-AB model, which  amplifies structure growth across all redshifts. 

Overall, the GP comparisons show that the current growth measurements are compatible with a broad class of cosmological scenarios. 
 %tends to favor models with slightly weaker growth than the standard $\Lambda$CDM prediction, although 
The reconstructed evolution from current data, producing the uncertainties shown in Figure~\ref{fig:sigma8_gm}, do not allow a statistically significant preference for any specific model.

The comparison presented in this subsection focuses on the cosmic evolution of the observables. 
We leave for the following subsection the statistical performance of the models, including the goodness-of-fit indicators and information criteria derived from the MCMC analysis. 

\begin{table*}[t]
\caption{Best-fit cosmological parameters obtained from the MCMC 
analysis for the modified gravity models.}
\centering
\begin{tabular}{lcccc}
\hline\hline
Model & $H_0$ & $\Omega_{m0}$ & $\sigma_8$ & Model parameter \\
\hline
$\Lambda$CDM &
\hspace{0.3cm} $69.98^{+6.91}_{-6.81}$ &
\hspace{0.3cm} $0.2767^{+0.0330}_{-0.0304}$ &
\hspace{0.3cm} $0.7716^{+0.0220}_{-0.0218}$ & -- \\[0.15cm]
Starobinsky ($n=1$) &
\hspace{0.3cm} $70.34^{+6.62}_{-6.91}$ &
\hspace{0.3cm} $0.3379^{+0.0994}_{-0.0997}$ &
\hspace{0.3cm} $0.7802^{+0.0288}_{-0.0317}$ &
\hspace{0.3cm} $\lambda_S=0.802^{+0.474}_{-0.482}$ \\[0.15cm]
Starobinsky ($n=2$) &
\hspace{0.3cm} $70.10^{+6.82}_{-6.71}$ &
\hspace{0.3cm} $0.3044^{+0.1342}_{-0.1387}$ &
\hspace{0.3cm} $0.7711^{+0.0206}_{-0.0206}$ &
\hspace{0.3cm} $\lambda_S=1.295^{+0.506}_{-0.512}$ \\[0.15cm]
Hu-Sawicki ($n=1$) &
\hspace{0.3cm} $70.00^{+6.82}_{-6.77}$ &
\hspace{0.3cm} $0.2709^{+0.0373}_{-0.0336}$ &
\hspace{0.3cm} $0.7719^{+0.0221}_{-0.0221}$ &
\hspace{0.3cm} $c_2=109.94^{+57.91}_{-64.31}$ \\[0.15cm]
Hu-Sawicki ($n=2$) &
\hspace{0.3cm} $69.77^{+6.87}_{-6.69}$ &
\hspace{0.3cm} $0.3007^{+0.0502}_{-0.0405}$ &
\hspace{0.3cm} $0.7706^{+0.0223}_{-0.0215}$ &
\hspace{0.3cm} $c_2=120.03^{+56.33}_{-71.44}$ \\[0.15cm]
$R^2$-AB &
\hspace{0.3cm} $67.47^{+5.13}_{-5.14}$ &
\hspace{0.3cm} $0.2051^{+0.0285}_{-0.0240}$ &
\hspace{0.3cm} $0.7723^{+0.0218}_{-0.0219}$ &
\hspace{0.3cm} $b=1.626^{+1.268}_{-0.870}$ \\
\hline\hline
\end{tabular}
\label{tab:MG_parameters}
\end{table*}

%\begin{table*}[htb]
%\centering
%\caption{Best-fit cosmological parameters obtained from the MCMC analysis for the modified gravity models.}
%\begin{tabular}{lcccc}
%\hline
%Model & \hspace{0.5cm}$H_0$ & \hspace{0.5cm}$\Omega_{m0}$ & \hspace{0.5cm}$\sigma_8$ & \hspace{0.5cm}Model parameter \\
%\hline
%$\Lambda$CDM &
%\hspace{0.5cm}$69.98^{+6.91}_{-6.81}$ &
%\hspace{0.5cm}$0.2767^{+0.0330}_{-0.0304}$ &
%\hspace{0.5cm}$0.7716^{+0.0220}_{-0.0218}$ &
%\hspace{0.5cm}-- \\

%Starobinsky ($n=1$) &
%\hspace{0.5cm}$70.34^{+6.62}_{-6.91}$ &
%\hspace{0.5cm}$0.3379^{+0.0994}_{-0.0997}$ &
%\hspace{0.5cm}$0.7802^{+0.0288}_{-0.0317}$ &
%\hspace{0.5cm}$\lambda_S=0.802^{+0.474}_{-0.482}$ \\

%Starobinsky ($n=2$) &
%\hspace{0.5cm}$70.10^{+6.82}_{-6.71}$ &
%\hspace{0.5cm}$0.3044^{+0.1342}_{-0.1387}$ &
%\hspace{0.5cm}$0.7711^{+0.0206}_{-0.0206}$ &
%\hspace{0.5cm}$\lambda_S=1.295^{+0.506}_{-0.512}$ \\

%5Hu-Sawicki ($n=1$) &
%\hspace{0.5cm}$70.00^{+6.82}_{-6.77}$ &
%\hspace{0.5cm}$0.2709^{+0.0373}_{-0.0336}$ &
%\hspace{0.5cm}$0.7719^{+0.0221}_{-0.0221}$ &
%\hspace{0.5cm}$c_2=109.94^{+57.91}_{-64.31}$ \\

%Hu-Sawicki ($n=2$) &%
%\hspace{0.5cm}$69.77^{+6.87}_{-6.69}$ &
%\hspace{0.5cm}$0.3007^{+0.0502}_{-0.0405}$ &
%\hspace{0.5cm}$0.7706^{+0.0223}_{-0.0215}$ &
%\hspace{0.5cm}$c_2=120.03^{+56.33}_{-71.44}$ \\

%$R^2$-AB &
%\hspace{0.5cm}$67.47^{+5.13}_{-5.14}$ &
%\hspace{0.5cm}$0.2051^{+0.0285}_{-0.0240}$ &
%\hspace{0.5cm}$0.7723^{+0.0218}_{-0.0219}$ &
%\hspace{0.5cm}$b=1.626^{+1.268}_{-0.870}$ \\

%\hline
%\end{tabular}
%\label{tab:MG_parameters}
%\end{table*}

\begin{table}[ht]
\caption{Goodness-of-fit statistics for the modified gravity 
models, obtained from the joint analysis combining the $f(z)$ 
and $\sigma_8(z)$ datasets.}
\centering
\begin{tabular}{lcccc}
\hline\hline
Model & $\chi^2_{\min}$ & $\chi^2_\nu$ & AIC & BIC \\
\hline
$\Lambda$CDM & 11.64 & 0.51 & 17.64 & 21.41 \\
Starobinsky ($n=1$) & 9.25  & 0.42 & 17.25 & 22.28 \\
Starobinsky ($n=2$) & 11.36 & 0.52 & 19.36 & 24.39 \\
Hu-Sawicki ($n=1$)  & 11.21 & 0.51 & 19.21 & 24.24 \\
Hu-Sawicki ($n=2$)  & 11.44 & 0.52 & 19.44 & 24.47 \\
$R^2$-AB            & 10.84 & 0.49 & 18.84 & 23.87 \\
\hline\hline
\end{tabular}
\label{tab:MG_statistics}
\end{table}

%\begin{table}[ht]
%\centering
%\caption{Goodness-of-fit statistics for the modified gravity models, obtained from the joint analysis combining the $f(z)$ and $\sigma_8(z)$ datasets.}

%\begin{tabular}{lcccc}
%\hline
%Model & \hspace{0.3cm}$\chi^2_{\min}$ & \hspace{0.3cm}$\chi^2_\nu$ & \hspace{0.3cm}AIC & \hspace{0.3cm}BIC\\
%\hline
%Starobinsky ($n=1$) & \hspace{0.3cm}9.25 & \hspace{0.3cm}0.42 & \hspace{0.3cm}17.25 & \hspace{0.3cm}22.43\\
%Starobinsky ($n=2$) & \hspace{0.3cm}11.36 & \hspace{0.3cm}0.52 & \hspace{0.3cm}19.36 & \hspace{0.3cm}24.54\\
%Hu-Sawicki ($n=1$) & \hspace{0.3cm}11.21 & \hspace{0.3cm}0.51 & \hspace{0.3cm}19.21 & \hspace{0.3cm}24.40\\
%Hu-Sawicki ($n=2$) & \hspace{0.3cm}11.44 & \hspace{0.3cm}0.52 & \hspace{0.3cm}19.44 & \hspace{0.3cm}24.63\\
%$R^2$-AB & \hspace{0.3cm}10.84 & \hspace{0.3cm}0.49 & \hspace{0.3cm}18.84 & \hspace{0.3cm}24.02\\
%\hline
%\end{tabular}
%\label{tab:MG_statistics}
%\end{table}

%%%%%%%%%%%%%%%%%%%%%%%%%%%%%%%%%%%%%%%%%%%%%%%%%%%%%%%%5
\subsection{Model comparison}
%%%%%%%%%%%%%%%%%%%%%%%%%%%%%%%%%%%%%%%%%%%%%%%%%%%%%%%%%%%%%%%
The statistical performance of the models is evaluated through the minimum chi-square, reduced chi-square, Akaike information criterion (AIC), and Bayesian information criterion (BIC), summarized in Table~\ref{tab:MG_statistics}. 
These quantities provide a quantitative assessment of how well each $F(R)$ model reproduces the current growth-rate of cosmic structures and clustering-amplitude measurements, while accounting for the additional parameters introduced by each scenario. 
All models present reduced chi-square values smaller than unity.  This behavior is not necessarily indicative of overfitting, since the present growth datasets contain a limited number of measurements and relatively large uncertainties, particularly for the ones used here. Therefore, the low values of $\chi^2_\nu$ mainly reflect the current constraining power of the data rather than necessarily indicating overfitting.

%Lower AIC and BIC values correspond to models that best balance goodness of fit and simplicity.
%indicating the best agreement with the growth data. 

Among the modified gravity scenarios, the Starobinsky $n=1$ model provides the lowest $\chi^2_{\rm min}$, AIC, and BIC values. It also yields the lowest reduced chi-square value, $\chi^2_\nu=0.42$. 
However, this result must be interpreted with caution. 
As presented in Appendix~\ref{altenative-models}, a stable de Sitter fixed point for this model requires $\lambda_S \geq 1.54$~\cite{DeFelice2010,Bessa2021}. 
The best-fit value obtained here, $\lambda_S = 0.802^{+0.474}_{-0.482}$, together with its full $1\sigma$ credible interval, lies entirely below this threshold. 
This indicates that the statistically preferred point in parameter space does not support a stable late-time de Sitter attractor, illustrating --within the very model favored by standard goodness-of-fit criterion-- the same tension between statistical performance and physical viability discussed for the $R^2$-AB model in the following subsections.

Overall, the reduced chi-square values for all models remain close, with $0.42\lesssim\chi^2_\nu\lesssim0.52$, showing that the different scenarios provide statistically comparable descriptions of the observed growth history. When comparing the MG models among themselves, the information criteria show only mild differences. 
Taking the Starobinsky $n=1$ model as reference, the alternative $F(R)$ scenarios present small variations, with $\Delta{\rm AIC}$ and $\Delta{\rm BIC}$ ranging from $1.59$ to $2.19$. The $R^2$-AB model exhibits the closest statistical performance, with $\Delta{\rm AIC}=\Delta{\rm BIC}=1.59$, followed by the Hu-Sawicki $n=1$ model ($\Delta{\rm AIC}=1.96$, $\Delta{\rm BIC}=1.97$), the Starobinsky $n=2$ model ($\Delta{\rm AIC}=\Delta{\rm BIC}=2.11$), and the Hu-Sawicki $n=2$ model ($\Delta{\rm AIC}=2.19$, $\Delta{\rm BIC}=2.20$). 
Therefore, according to the standard interpretation of information criteria, 
because the differences remain small, we conclude that none of the considered $F(R)$ scenarios is statistically preferred over the others.

We further compare the MG scenarios with the standard $\Lambda$CDM scenario. Relative to $\Lambda$CDM, the Starobinsky $n=1$ model provides a slightly lower AIC value, corresponding to $\Delta{\rm AIC}=-0.39$, while its BIC difference is positive, $\Delta{\rm BIC}=0.87$. 
The remaining scenarios do not improve the AIC with respect to $\Lambda$CDM, with $\Delta{\rm AIC}$ values ranging from $1.20$ for the $R^2$-AB model to $1.80$ for the Hu-Sawicki $n=2$ model. Their BIC differences are also positive, varying from $2.46$ to $3.06$. These results indicate that, although some MG models can provide a slightly better fit to the growth data, the improvement is not statistically significant once the additional model complexity is taken into account. Therefore, the current growth data do not provide strong evidence for a departure from General Relativity in favor of modified gravity over the standard $\Lambda$CDM scenario.

The $R^2$-AB model presents the second lowest AIC and BIC values among the MG scenarios, while the remaining models exhibit similar information criteria. These results indicate that several $F(R)$ models can reproduce the observed growth evolution with comparable accuracy. However, information criteria alone do not determine the physical viability of the models, since scenarios with similar statistical performance may predict significantly different modifications of the gravitational interaction. For this reason, we complement the statistical analysis with an investigation of the effective gravitational coupling, which directly quantifies deviations from GR at cosmological scales.

%%%%%%%%%%%%%%%%%%%%%%%%%%%%%%%%%%%%%%%%%%%%%%%%%%%%%%%%%5
\subsection{Effective gravitational coupling}
%%%%%%%%%%%%%%%%%%%%%%%%%%%%%%%%%%%%%%%%%%%%%%%%%%%%%%%

Beyond the statistical diagnostics discussed above, it is also interesting to investigate the dynamical behavior of gravity predicted by each scenario. For this purpose, we analyze the effective gravitational coupling, $\mu(z)$, introduced in Section \ref{basics}. Since GR predicts $\mu(z)=\text{const.}=1$ at all redshifts, departures from unity provide a direct measure of modifications to  the gravitational interaction.

\begin{figure}[h!]
\centering
\includegraphics[width=0.5\textwidth]{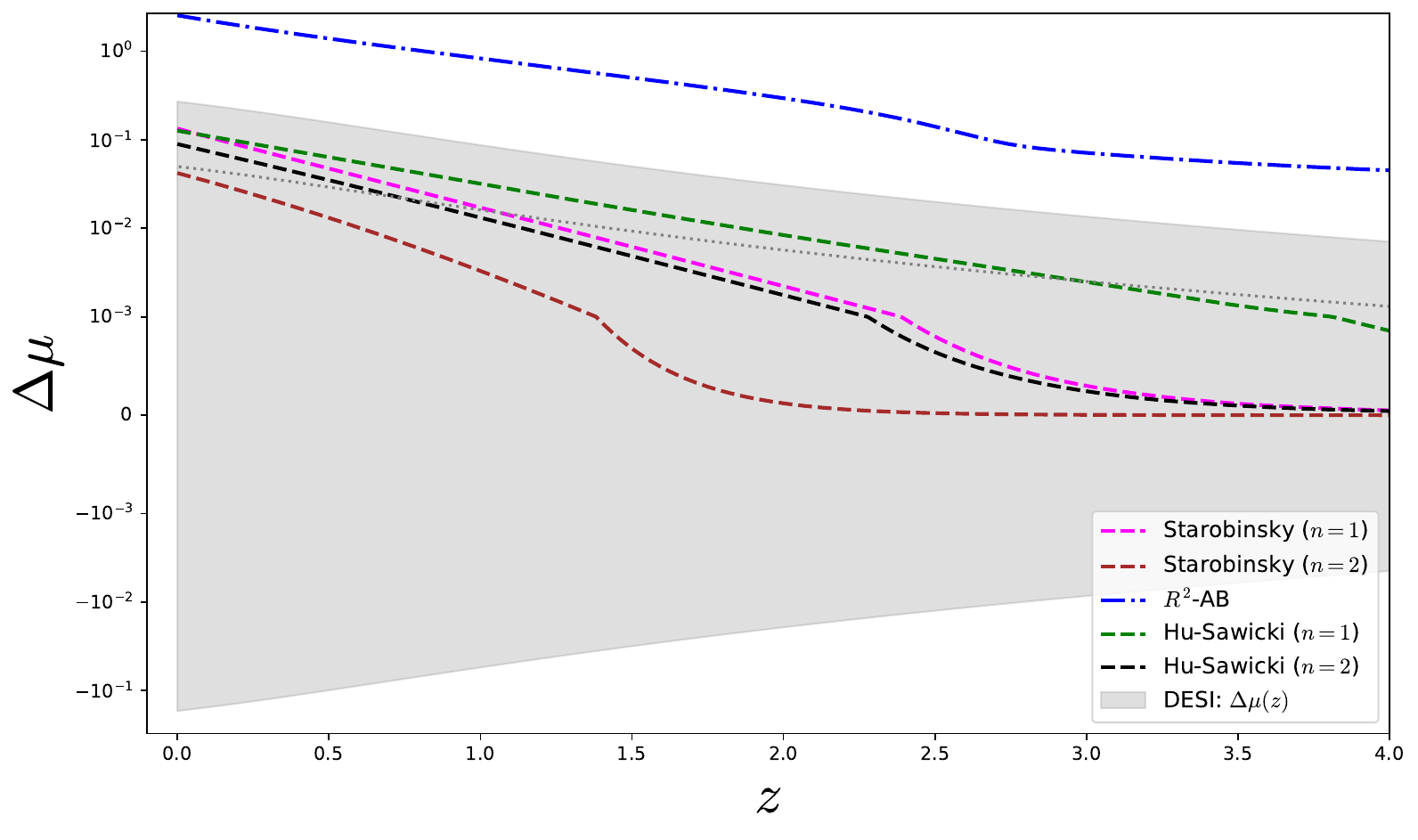}
\caption{Redshift evolution of $\Delta\mu(z) \equiv \mu(z,\bar{k}) - 1$ for the analyzed cosmological models. 
The shaded region represents the $1\sigma$ DESI constraint~\cite{Ishak2024}, $\Delta\mu(z) = (0.05 \pm 0.22)\,\Omega_{\rm DE}(z)/\Omega_\Lambda$, which narrows with redshift as dark energy becomes subdominant.}
\label{fig:muz}
\end{figure}

Figure~\ref{fig:muz} presents the redshift 
evolution of $\Delta \mu(z) = \mu(z,\bar{k}) - 1$ for the MG scenarios considered in this work.  
As expected for viable $F(R)$ theories, all models approach the GR limit at high redshift, i.e. $\Delta \mu(z)=0$, where the curvature is large and the modifications become negligible. The deviations therefore emerge predominantly at late times, when the scalar degree of freedom associated with
$F(R)$ gravity becomes dynamically relevant. Such behavior is consistent with MG frameworks capable of explaining the late time accelerated expansion without introducing a cosmological constant \cite{andrade, Salvatelli_2016}. 

The Starobinsky and Hu-Sawicki models exhibit only moderate departures from GR, with $\mu(z)$ remaining close to unity over the entire redshift range. 
Although both families predict a mild enhancement of the effective gravitational coupling at low redshift, their redshift evolution is not identical. 
The Starobinsky models converge more rapidly toward the GR limit as redshift increases, whereas the Hu-Sawicki models retain a slightly enhanced coupling over a broader redshift interval before asymptotically approaching to 
$\Delta \mu(z) = 0$. 
A markedly different behavior is found for the $R^2$-AB theory, which predicts $\Delta \mu_0 \sim 2.5$, corresponding to an effective gravitational coupling more than three times larger than Newton's gravitational constant at current times. 
We note that this large present-day coupling is physically associated with the anomalously low $\Omega_{m0} \simeq 0.205$ preferred by the MCMC, reflecting a compensation between stronger 
gravitational coupling and reduced matter content that will be 
further discussed in Section~\ref{s8}. 
Moreover, Figure~\ref{fig:muz} shows that this enhancement persists over a wide redshift interval, remaining significantly above the GR prediction up to $z\sim2-3$ before gradually converging toward zero. Such a strong and long-lasting amplification of gravity is difficult to reconcile with the expected behavior of viable modified gravity models, which generally require only small deviations from GR on cosmological scales while simultaneously satisfying local gravity constraints.

\begin{figure}[ht]
\centering    
\includegraphics[width=0.5\textwidth]{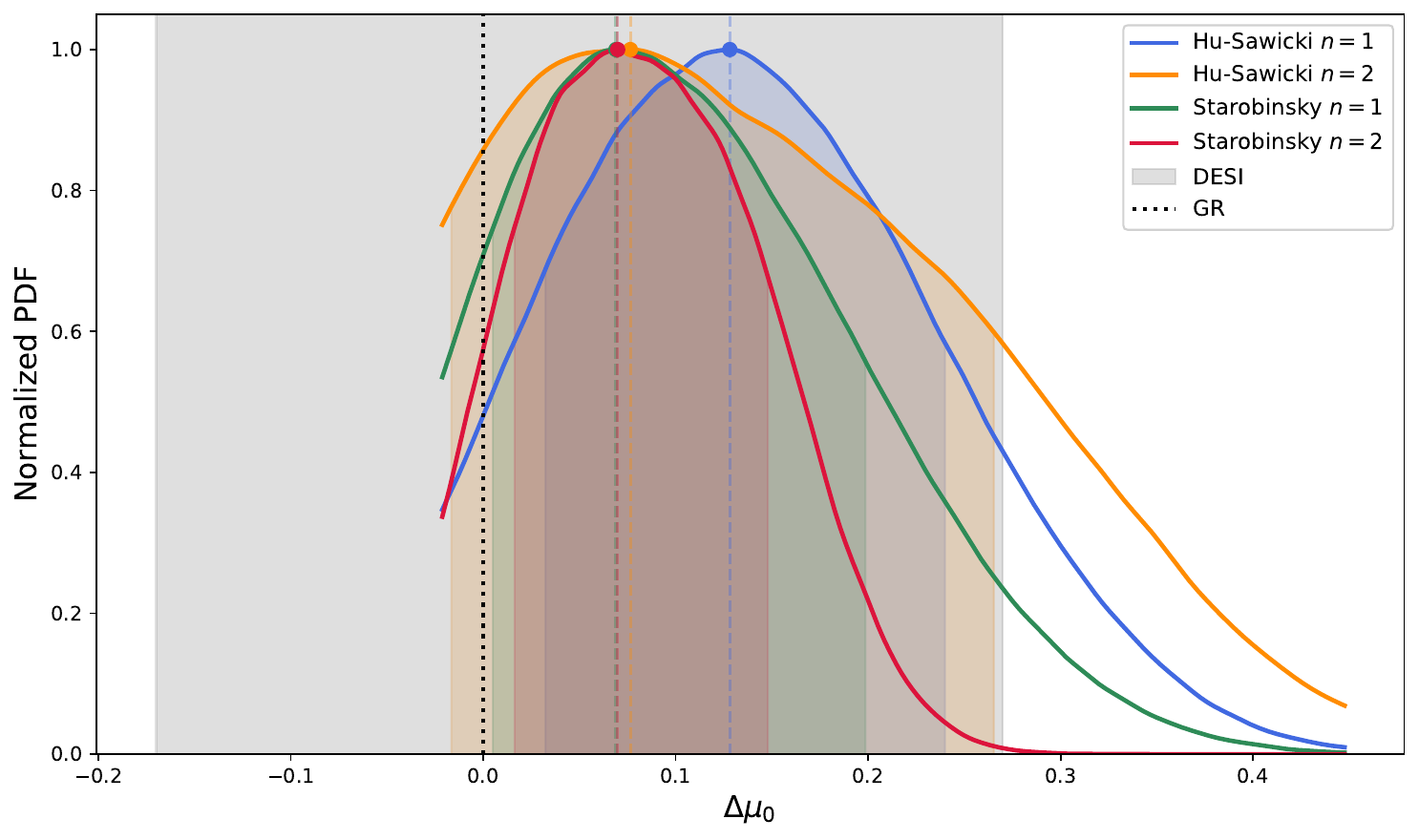}
\caption{
Normalized PDFs of the present-day effective 
gravitational coupling deviation $\Delta\mu_0 \equiv \Delta\mu(z=0)$, obtained from 1000 MCMC posterior samples for the Hu-Sawicki and Starobinsky models with $n=1$ and $n=2$. The dashed vertical 
lines indicate the best-fit values reported in 
Table~\ref{tab:MG_parameters}. 
The gray-shaded region corresponds 
to the $1 \sigma$ DESI constraint, $\mu_0 - 1 = 0.05 \pm 
0.22$~\cite{Ishak2024}. The red dotted vertical line marks 
the GR prediction, $\Delta\mu_0 = 0$.}%PDFs of the present-day gravitational coupling, $\mu_0 = G_{\rm eff}(z=0) / G_N$, obtained from 1000 MCMC samples for the Hu-Sawicki and Starobinsky models with $n=1$ and $n=2$. The dashed vertical lines indicate the best-fit values reported in 
%\textcolor{red}{Table~\ref{tab:geff}} 
%\textcolor{red}{@Wilian: a tabela com os valores de $G_{eff}$ foi removida?}, while the shaded region corresponds to the constraint at the $\pm1\sigma$ confidence level. The red dotted vertical line marks the GR prediction, $\mu_0 = 1$.}    
\label{fig:pdf_mu0}
\end{figure}

Figure~\ref{fig:pdf_mu0} provides a complementary statistical characterization through the normalized posterior probability density functions (PDFs) of the 
present day value $\Delta \mu_0$. 
The PDFs closely reflect the qualitative behavior observed in Figure~\ref{fig:muz}.  
The Hu-Sawicki and Starobinsky models retain a substantial overlap with the GR prediction and the observational error 
of $\Delta \mu_0$ (gray shadows) at 1$\sigma$. 
By contrast, the PDFs show that the $R^2$-AB model favors substantially larger present day values of 
$\Delta \mu_0$ (not shown). 
This result reinforces the conclusion that, despite providing a competitive statistical fit to the growth data, the model requires a level of gravitational amplification considerably stronger than that predicted by the other viable $F(R)$ scenarios.

Taken together, Figures~\ref{fig:muz} and \ref{fig:pdf_mu0} illustrate an important aspect of modified gravity analyses. Models with comparable goodness-of-fit statistics may predict substantially different gravitational dynamics. While the information criteria discussed in the previous subsection reveal only small statistical differences among the $F(R)$ models, the effective gravitational coupling provides additional physical discrimination. In particular, the Hu-Sawicki and Starobinsky models remain close to the GR limit throughout cosmic evolution, whereas the $R^2$-AB model predicts a much stronger modification of gravity. Consequently, the evolution of $\mu(z)$, together with its posterior distribution, constitutes an important complementary criterion for assessing the physical viability of MG scenarios beyond purely statistical goodness-of-fit indicators.

%%%%%%%%%%%%%%%%%%%%%%%%%%%%%%%%%%%%%%%%%%%%%%%%5
\subsection{Comparative analysis of $S_8$} \label{s8}
%%%%%%%%%%%%%%%%%%%%%%%%%%%%%%%%%%%%%%%%%%%%%%%%%%
As a complementary consistency test, we compare the values of the derived parameter $S_8\equiv \sigma_8\sqrt{{\Omega_{m0}}/{0.3}}$ with the \textit{Planck} 2018 reference measurement~\cite{Planck2018}, $S_8^{\rm Planck}=0.832 \pm 0.013$. Since $S_8$ combines the present day matter density and the clustering amplitude, it provides a convenient summary statistic for assessing the growth of cosmic structures and has become one of the most widely used quantities for comparing cosmological models. 

\begin{table}[h]
\caption{Derived $S_8$ values for the modified gravity models 
and their statistical tension with respect to the Planck 2018 
measurement ($S_8 = 0.832 \pm 0.013$). The quoted uncertainties 
were obtained through standard propagation of the asymmetric 
uncertainties in $\Omega_{m0}$ and $\sigma_8$.}
\centering
\begin{tabular}{lcc}
\hline\hline
Model & $S_8$ & \,\,\,Tension [$\sigma$] \\
\hline
Starobinsky ($n=1$) & \,\,\,$0.828^{+0.149}_{-0.161}$ & 0.03 \\[0.15cm]
Starobinsky ($n=2$) & \,\,\,$0.777^{+0.181}_{-0.219}$ & 0.28 \\[0.15cm]
Hu-Sawicki ($n=1$)  & \,\,\,$0.734^{+0.071}_{-0.067}$ & 1.41 \\[0.15cm]
Hu-Sawicki ($n=2$)  & \,\,\,$0.771^{+0.086}_{-0.074}$ & 0.76 \\[0.15cm]
$R^2$-AB            & \,\,\,$0.638^{+0.062}_{-0.056}$ & 3.20 \\
\hline\hline
\end{tabular}
\label{tab:S8}
\end{table}

%\begin{table}[h!]
%\caption{Derived $S_8$ values for the modified gravity models and their statistical tension with respect to the Planck 2018 measurement ($S_8=0.832\pm0.013$). The quoted uncertainties were obtained through standard propagation of the asymmetric uncertainties in $\Omega_{m0}$ and~$\sigma_8$.}

%\centering
%\begin{tabular}{lcc}
%\hline
%Model & \hspace{0.6cm}$S_8$ & \hspace{0.5cm}Tension [$\sigma$] \\
%\hline
%Starobinsky ($n=1$) & \hspace{0.7cm}$0.828^{+0.149}_{-0.161}$ & \hspace{0.5cm}0.03 \\
%Starobinsky ($n=2$) & \hspace{0.7cm}$0.777^{+0.181}_{-0.219}$ & \hspace{0.5cm}0.28 \\
%Hu-Sawicki ($n=1$) & \hspace{0.7cm}$0.734^{+0.071}_{-0.067}$ & \hspace{0.5cm}1.41 \\
%Hu-Sawicki ($n=2$) & \hspace{0.7cm}$0.771^{+0.086}_{-0.074}$ & \hspace{0.5cm}0.76 \\
%$R^2$-AB & \hspace{0.7cm}$0.638^{+0.062}_{-0.056}$ & \hspace{0.5cm}3.20 \\
%\hline
%\end{tabular}
%\label{tab:S8}
%\end{table}

The values of $S_8$ inferred for each MG model are listed in Table~\ref{tab:S8}, where the quoted uncertainties were obtained by propagating the asymmetric uncertainties of $\Omega_{m0}$ and $\sigma_8$. To quantify the agreement with the Planck determination, we compute the statistical tension
\begin{equation}
T = \frac{|S_8^{\rm model} - S_8^{\rm Planck}|}{\sigma_{\rm comb}}.
\end{equation}
where $\sigma_{\rm comb} \equiv \sqrt{\sigma^2_{\rm model} + \sigma^2_{\rm Planck}}$, 
$\sigma_{\rm Planck}=0.013$ is the uncertainty of the Planck measurement, and 
$\sigma_{\rm model}$ was taken as the propagated uncertainty in the direction of the Planck value (i.e., the upper or lower asymmetric uncertainty, depending on whether the model prediction lies below or above the Planck measurement). 

The results show that all models, with the exception of the $R^2$-AB scenario, are statistically consistent with the Planck determination within approximately $1.5\sigma$. In particular, the Starobinsky model with $n=1$ yields $S_8=0.828$, which is essentially indistinguishable from the Planck value. 
The Starobinsky ($n=2$) and Hu-Sawicki ($n=2$) models also remain fully compatible with the Planck constraint, exhibiting tensions well below the $1\sigma$ level.

The Hu-Sawicki model with $n=1$ predicts a slightly smaller value, $S_8=0.734$, resulting in a moderate tension of about $1.4\sigma$. Although this is the largest discrepancy among the viable $F(R)$ models, it remains well below the threshold required to claim a statistically significant disagreement with the Planck measurement. A markedly different behavior is found for the $R^2$-AB model. 
Its low matter density parameter ($\Omega_{m0}\simeq0.21$) leads to $S_8=0.638$, corresponding to a tension of approximately $3.2\sigma$ with respect to the Planck reference, a direct consequence of the low value of $\Omega_{m0}$ obtained in this model. The value inferred for $\sigma_8 \approx 0.77$ is broadly consistent with the other $F(R)$ scenarios considered here. 
This result is consistent with the strong enhancement of the effective gravitational coupling discussed in the previous subsection and provides independent evidence that this scenario is disfavored despite its competitive goodness-of-fit statistics.

Taken together, the goodness-of-fit analysis, the evolution of the effective gravitational coupling, and the $S_8$ comparison reveal a coherent picture. While several $F(R)$ models reproduce the current growth measurements with comparable statistical quality, only those predicting moderate deviations from GR remain simultaneously compatible with both the inferred gravitational dynamics and the Planck constraint on structure growth. In this context, the Starobinsky and Hu-Sawicki models emerge as the most physically plausible modified gravity scenarios within the present observational uncertainties, whereas the $R^2$-AB model is disfavored by its pronounced departures in both $\mu(z)$ and $S_8$.

%%%%%%%%%%%%%%%%%%%%%%%%%%%%%%%%%%%%%%%%%%%%%%%%%%%%%%%%%5
\section{Conclusions}\label{conclusions}
%%%%%%%%%%%%%%%%%%%%%%%%%%%%%%%%%%%%%%%%%%%%%%%%%%%%%%%%%%
In this work, we investigate alternative cosmological models that emerged as modifications of the GR theory, termed $F(R)$ MG cosmological models. 
For this study we perform a MCMC analysis of three representative $F(R)$ theories: Starobinsky, Hu–Sawicki, and $R^2$–AB, using current measurements of the  growth rate $f(z)$ and clustering amplitude $\sigma_8(z)$. 
The MCMC constraints were complemented by GP reconstructions, used  as a model-independent benchmark for the evolution of these growth observables. 
Our statistical analyses show that all investigated $F(R)$ models provide comparably good fits to the current growth data, with differences in the quantities $\chi^2$, $\text{AIC}$, and $\text{BIC}$ remaining too small to establish a statistically significant preference for any particular scenario. These results indicate that the present growth measurements alone do not possess sufficient discriminating power to distinguish among viable $F(R)$  models based exclusively on standard goodness-of-fit criteria.

To overcome this limitation, we complemented the statistical comparison with two physically motivated diagnostics: the reconstructed effective gravitational coupling, $\mu(z)$, and the derived cosmological parameter $S_8$. 
Unlike the information criteria, these quantities probe the gravitational interaction of each theory 
%dynamics predicted by 
and therefore provide an additional level of physical discrimination. 
%Although the Starobinsky $n=1$ model provides the best statistical fit, 
From these analyses, we conclude that all the viable $F(R)$ models considered here, with the exception of the $R^2$-AB model, remain compatible with the cosmological constraints on both the effective gravitational coupling, $\mu(z)$, and the derived parameter $S_8$. 
Therefore, the present growth data do not favor a single modified gravity scenario but instead indicate that several $F(R)$ models can reproduce the current observations while predicting only moderate departures from GR.

As a matter of fact, the $R^2$-AB model presents a qualitatively different behavior. 
Despite its competitive statistical performance, it predicts a substantially enhanced effective gravitational coupling together with a low value of $S_8$, leading to a tension of approximately $3.2\sigma$ with the Planck measurement. 
%illustrates 
This result reminds us that a satisfactory statistical fit does not necessarily imply physical viability, emphasizing the importance of complementing information criteria with diagnostics that are directly sensitive to the underlying gravitational dynamics.

Overall, our results show that current growth data remain consistent with viable $F(R)$ theories that predict only moderate departures from GR theory. 
Within the class of models investigated here, the  Hu-Sawicki and Starobinsky 
($n=2$) models provide the most balanced combination of statistical performance and physical consistency. 
More generally, our investigation demonstrates that combining MCMC parameter estimation with the reconstruction of the effective gravitational coupling constitutes a robust framework for testing modified gravity models with present and future large-scale growth structure observations. 
Future large and deep astronomical surveys will considerably improve the precision of growth measurements, making physically motivated diagnostics such as the effective gravitational coupling $\mu(z)$ increasingly important to distinguish between viable MG theories, that currently remain statistically indistinguishable using standard goodness-of-fit criteria alone.

\begin{acknowledgments}
%\noindent  
MVSF and FO thank CAPES for their fellowships. WSHR acknowledges FAPES and CNPq for partial financial support. 
FA thanks to Fundação de Amparo à Pesquisa do Estado do Rio de Janeiro (FAPERJ), Processo SEI-260003/001221/2025,  for the financial support.  
AB acknowledges a CNPq fellowship. 
\end{acknowledgments}

\bibliography{main}
\appendix
\section{Gaussian Processes} \label{appendixA}
To obtain a model-independent reconstruction of cosmological functions, we employ the GP method as developed by \cite{Seikel2012}. This non-parametric, fully Bayesian approach allows one to infer a continuous function from discrete data without assuming a specific parametrization for its evolution~\cite{Seikel13, Yang2015, Valent2018, Duan2015, Jesus2019, Oliveira2023}. In the present analysis, we apply this technique to reconstruct the redshift dependence of the growth rate of structure, $f(z)$, and the matter fluctuation amplitude, $\sigma_8(z)$, directly from observational data~\cite{Avila22a, Avila22b, Zhang2018, Blake2011}.

GP generalizes the Gaussian probability distribution to a distribution over functions~\cite{Vanhatalo2012}. The value of a function $q$ evaluated at a point $x$ is a Gaussian random variable with mean $\eta(x)$. 
Function values at different points, $x$ and $\tilde{x}$, are generally correlated, with the correlation described by a covariance function $k(x, \tilde{x})$
\begin{equation}
\eta(x) = \mathbb{E}[q(x)], \qquad
k(x, \tilde{x}) = \mathbb{E}[(q(x) - \eta(x))(q(\tilde{x}) - \eta(\tilde{x}))] \,,
\end{equation}
where $\mathbb{E}$ is the expected value for these functions. 

Thus, GP can be expressed as
\begin{equation}
q(x) \sim \mathcal{GP}\!\left(\eta(x), k(x, \tilde{x})\right),
\end{equation}
where $\eta(x)$ defines the mean of the process and $k(x, \tilde{x})$ encodes the correlations between function values at different points.

In this work, we adopt the squared exponential covariance function, which guarantees smoothness and differentiability of the reconstructed functions
\begin{equation}
k(x, \tilde{x}) = \sigma_f^2 \exp\!\left[-\frac{(x - \tilde{x})^2}{2 \ell^2}\right],
\end{equation}
where $\sigma_f$ and $\ell$ are hyperparameters controlling, respectively, the amplitude and correlation length (or smoothness scale) of the process. These hyperparameters are optimized via marginal likelihood maximization. The choice of the covariance function does not influence the reconstruction, as shown in recent works~\cite{Oliveira2023, Oliveira2025a, Hwang2023, Zhang2023}.

%This methodology enables the reconstruction of the function itself as well as its derivatives, since the derivative of  GP is also a GP. %Consequently, one can consistently compute physical quantities depending on $f'(z)$ or higher derivatives.

Following the reconstruction procedure described in~\cite{Seikel2012}, we apply the GP technique to the datasets of $f(z)$ and $\sigma_8(z)$, obtaining smooth, model-independent reconstructions of their redshift evolution with associated confidence regions. These reconstructions are subsequently employed for comparison with theoretical predictions of the cosmological models considered in this work.

\section{Alternative Cosmological Models}\label{altenative-models}

Here, we briefly describe the alternative cosmological models considered in this work. For completeness, we present  GR-based extensions of the flat-$\Lambda$CDM model, whose 
results are discussed in Appendix~C. The $F(R)$ modified gravity theories, which are the primary focus of this work, are presented in the following subsection.

%%------------------------------------------
\subsection{Modified Gravity models: $F(R)$} \label{modelsfr}

The modified gravity theory is a prominent extension of GR where the Einstein-Hilbert Lagrangian, $\mathbf{L} \propto R$, with $R$ being the spacetime curvature, is replaced by an arbitrary function of the Ricci scalar, $\mathbf{L} \propto F(R)$~\cite{Capozziello2011, Clifton2011, DeFelice2010, Sotiriou2008, Ribeiro23, Oliveira2025a}. 
These $F(R)$ models aim to explain cosmic acceleration by modifying the gravitational law on large scales, potentially eliminating the need for an exotic dark energy component~\cite{Clifton2011, Tsujikawa2008, Capozziello2002}. 
Viable $F(R)$ models must satisfy stringent constraints, such as recovering GR in high-curvature environments and ensuring the stability of the de Sitter vacuum.

Among the most studied $F(R)$ models is the Hu-Sawicki model, designed to satisfy viability constraints and mimic $\Lambda$CDM behavior. 
Its functional form is given by \cite{Hu2007}
\begin{equation}
F_{\text{HS}}(R) = R - m_0^2 \frac{c_1 (R/m_0^2)^n}{c_2 (R/m_0^2)^n + 1}\,,
\label{eq:fR_HS}
\end{equation}
where $m_0^2 \equiv H_0^2 \,\Omega_{m0}$, with $c_1, c_2, n$ denoting the model parameters. The general relativistic limit is recovered when $c_1/c_2 \to 0$ while keeping $c_1/c_2$ fixed. In this limit, the effective cosmological constant is expressed as 
\begin{equation}
\Lambda = \frac{m_0^2 c_1}{2c_2}\,.
\end{equation}
Given that $\Lambda = 3H_0^2(1-\Omega_{m0})$, the parameters are related through
\begin{equation}
    c_1 = 6 \,c_2 \,\frac{1 - \Omega_{m0}}{\Omega_{m0}}\,,
\end{equation} 
which implies that the model has two free parameters: $n$ and $c_2$. The exponent $n$ controls the deviation from GR, with the cases $n=1$ and $n=2$ being frequently investigated as they represent the simplest non-trivial power-law modifications, offering distinct predictions for the growth of cosmic structures~\cite{Bessa2021, Nunes2017}.

Another interesting MG model is the Starobinsky model \cite{Starobinsky2007}, which is a generalized, viable form of the original $R+R^2$ model
\begin{equation}
F_{\text{S}}(R) = R + \lambda_S R_S \left[ \left( 1 + \frac{R^2}{R_S^2} \right)^{-n} - 1 \right] \,,
\label{eq:fR_SI}
\end{equation}
where $R_S$, $\lambda_S$, and $n > 0$ represent the model parameters. 
In the high-curvature limit, $R \gg R_S$, the model approaches an effective cosmological constant given by
\begin{equation}
\Lambda \equiv \frac{\lambda_S R_S}{2} \,.
\end{equation} 
The current curvature scale $R_S$ is connected to the parameter $\lambda_s$ through the relation
\begin{equation}
R_S = \frac{6H_0^2 (1 - \Omega_{m0})}{\lambda_S}.
\end{equation}
The parameter $n$ is also related to $\lambda_S$, which leads to lower bounds on $\lambda_S$ for a stable de Sitter solution. 
The requirement of a stable de~Sitter solution imposes 
lower bounds on $\lambda_S$, with $(n,\,\lambda_{S,\min}) 
\simeq (1,\,1.54),\,(2,\,0.94)$~\cite{DeFelice2010,Bessa2021}. 
In this work, we consider $n = 1$ and $n = 2$, leaving $\lambda_S$ as the only free parameter~\cite{Bessa2021, Motohashi2009, Kopp2013}.

Finally, the $R^2$-corrected Appleby-Battye ($R^2$-AB) model \cite{Appleby2007, Appleby2009} is an $F(R)$ model constructed to reproduce the $\Lambda$CDM expansion history while satisfying local gravity constraints. 
The functional form, governed by two free parameters, $\epsilon_{AB}$ and $b$, is 
\begin{equation}
F_{\text{AB}}(R) = \frac{R}{2} + \frac{\epsilon_{AB}}{2} \ln \left[\frac{\cosh(R/\epsilon_{AB}-b)}{\cosh b} + \frac{R^2}{6M^2} \right] ,
\label{eq:fR_AB}
\end{equation}
where $\epsilon_{AB}$ and $b$ are the model parameters, related by
\begin{equation}
\epsilon_{AB} = \frac{R_{\text{vac}}}{b + \ln ( 2 \cosh b )} \,,
\end{equation}
where $R_{\text{vac}} \equiv 12H_0^2$ denotes the vacuum scalar curvature. 
To account for the current cosmic acceleration, the model requires the condition 
$b \geq 1.6$ \cite{Appleby2009, Ribeiro23,motohashi}. 
After enforcing the de Sitter vacuum constraints, this framework stands out for 
introducing only one additional free parameter beyond the standard flat-$\Lambda$CDM model.
\begin{table*}[hbt]
\caption{Best-fit cosmological parameters, goodness-of-fit 
statistics, and derived $S_8$ values for the GR-based models, 
obtained from the joint $f(z)$ and $\sigma_8(z)$ analysis. 
Tensions are computed with respect to the Planck 2018 
reference measurement ($S_8 = 0.832 \pm 0.013$).}
\centering
\begin{tabular}{lccccccccc c}
\hline\hline
Model & $H_0$ & $\Omega_{m0}$ & $\sigma_8$ & 
Model parameter & \,\,$\chi^2_{\min}$\,\, & \,\,$\chi^2_\nu$\,\, & \,\,AIC\,\, & \,\,BIC\,\, &
\,\,$S_8$\,\, & Tension [$\sigma$] \\
\hline
flat-$\Lambda$CDM &
$69.98^{+6.91}_{-6.81}$ &
$0.2767^{+0.0330}_{-0.0304}$ &
$0.7716^{+0.0220}_{-0.0218}$ & -- &
\,\,11.64\,\, & \,\,0.51\,\, & \,\,17.64\,\, & \,\,21.41\,\, &
\,\,$0.741^{+0.065}_{-0.062}$\,\, & 1.41 \\[0.15cm]
$\omega$CDM &
$70.02^{+6.76}_{-6.85}$ &
$0.3133^{+0.1058}_{-0.1059}$ &
$0.7708^{+0.0218}_{-0.0214}$ &
$\omega=-0.885^{+0.265}_{-0.341}$ &
11.67 & 0.53 & 19.67 & 24.70 &
$0.788^{+0.149}_{-0.165}$ & 0.28 \\[0.15cm]
$\omega_0\omega_a$CDM &
$70.02^{+6.79}_{-6.87}$ &
$0.2949^{+0.0944}_{-0.0967}$ &
$0.7712^{+0.0216}_{-0.0218}$ &
$\omega_0=-0.891^{+0.388}_{-0.510}$ &
11.83 & 0.56 & 21.83 & 28.12&
$0.765^{+0.139}_{-0.155}$ & 0.45 \\
& & & &
$\omega_a=-0.212^{+1.243}_{-1.063}$ &
& & & & & \\[0.15cm]
$\Omega_k$CDM &
$70.28^{+6.68}_{-6.98}$ &
$0.3569^{+0.0499}_{-0.0477}$ &
$0.7689^{+0.0218}_{-0.0217}$ &
$\Omega_k=0.347^{+0.112}_{-0.212}$ &
16.55 & 0.75 & 24.55 & 29.58 &
$0.839^{+0.082}_{-0.080}$ & 0.04 \\
\hline\hline
\end{tabular}
\label{tab:GR_combined}
\end{table*}

%%------------------------------------------
\subsection{General Relativity-based Models}

The $\omega$CDM model treats the dark energy equation of 
state as a free constant $\omega \neq -1$, with
\begin{equation}
H(a) = H_0\sqrt{\Omega_{m0}a^{-3} + 
\Omega_{\Lambda 0}a^{-3(1+\omega)}}.
\tag{B1}
\end{equation}

The CPL parametrization allows a time-varying equation 
of state $\omega(a) = \omega_0 + \omega_a(1-a)$, giving
\begin{equation}
H(a) = H_0\sqrt{\Omega_{m0}a^{-3} + 
\Omega_{\Lambda 0}a^{-3(1+\omega_0+\omega_a)}e^{-3\omega_a(a-1)}}.
\tag{B2}
\end{equation}

The $\Omega_k$CDM model introduces spatial curvature 
through $\Omega_{k0}$:
\begin{equation}
H(a) = H_0\sqrt{\Omega_{m0}a^{-3} + 
\Omega_{k0}a^{-2} + \Omega_{\Lambda 0}},
\tag{B3}
\end{equation}
with $\Omega_{m0} + \Omega_{k0} + \Omega_{\Lambda 0} = 1$.

\section{Analyses of $\Lambda$CDM-type models}

For completeness, we present the GP reconstructions and statistical analyses for the GR-based extensions of the flat-$\Lambda$CDM model introduced in Appendix~B\,1. 
These results serve as a reference for comparison with the $F(R)$ scenarios 
discussed in the main text.

\begin{figure}[h]
\centering
\includegraphics[width=0.5\textwidth]{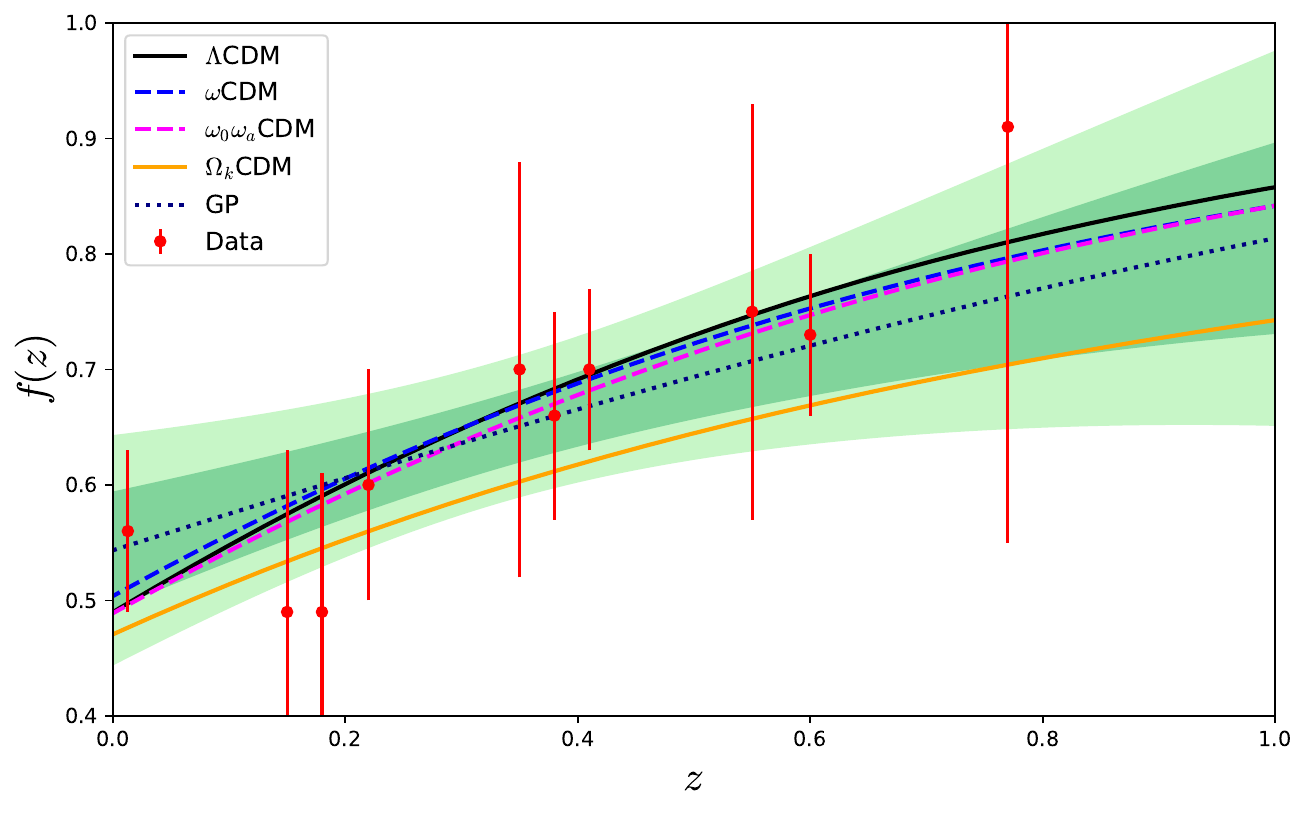}
\caption{Comparison among GP reconstruction of $f(z)$, 
for the flat-$\Lambda$CDM and $\Lambda$CDM-type models.}
\label{fig:fz_rg}
\end{figure}

\begin{figure}[H]
\centering
\includegraphics[width=0.5\textwidth]{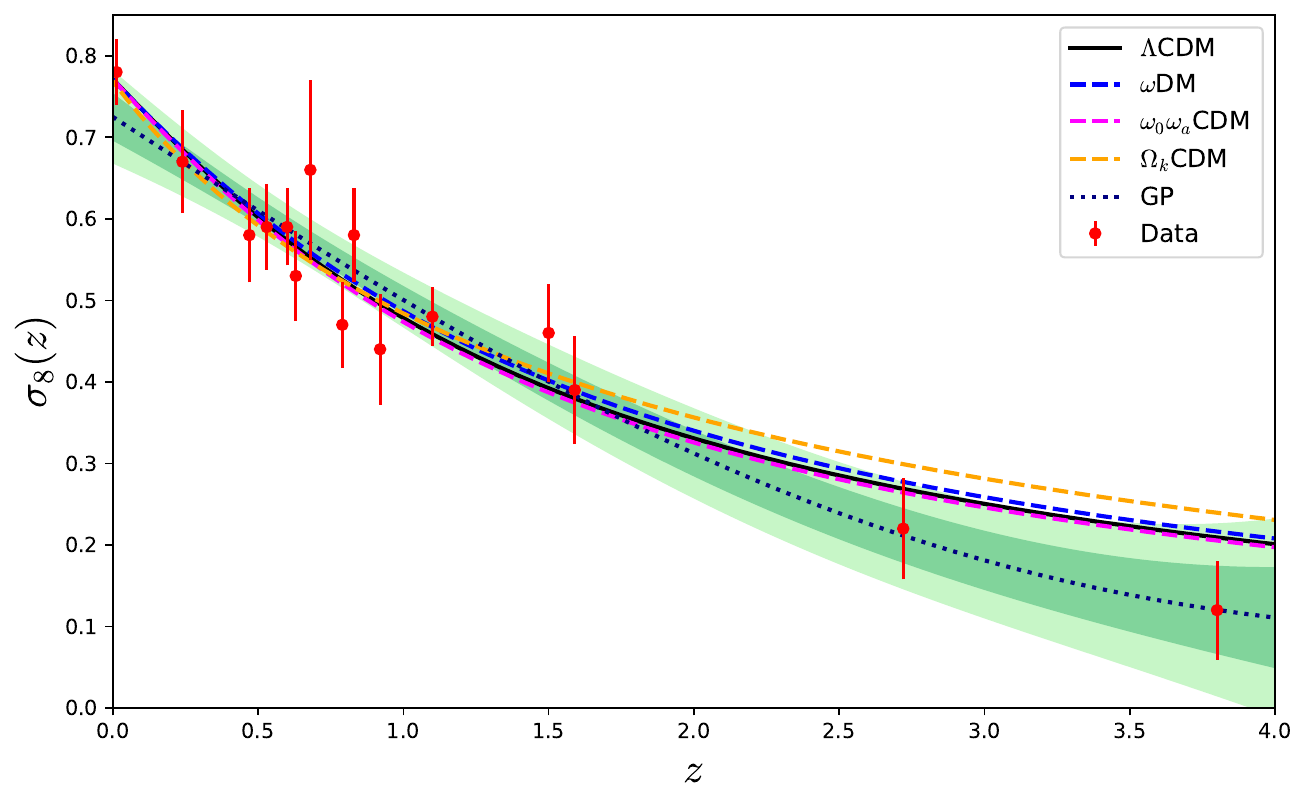}
\caption{Comparison among GP reconstruction of $\sigma_8(z)$, 
for the flat-$\Lambda$CDM and $\Lambda$CDM-type models.}
\label{fig:sigma8_rg}
\end{figure}

Figure~\ref{fig:fz_rg}  compares the best-fit predictions 
of the GR-based models with the GP reconstruction of $f(z)$. 
All models remain within the GP $2\sigma$ confidence region, 
with the exception of the $\Omega_k$CDM model, which consistently presents lower values of $f(z)$ compared to all other models and to the GP reconstruction. Figure~\ref{fig:sigma8_rg} presents the corresponding comparison for $\sigma_8(z)$. 
All models show concordance with the GP reconstruction at the $2\sigma$ confidence level, except the $\Omega_k$CDM model, which deviates significantly at high redshifts $z \gtrsim 2.5$.

The MCMC best-fit parameters for the GR-based models are summarized in Table~\ref{tab:GR_combined}. 
The inferred values of $H_0$, $\Omega_{m0}$, and $\sigma_8$ are broadly consistent among all scenarios and with those obtained for 
the $F(R)$ models (see Table~\ref{tab:MG_parameters}), confirming 
that the current growth data place similar constraints on the background cosmology regardless of the model considered. 
The derived $S_8$ values and their tensions with respect to the Planck 2018 measurement are summarized in Table~\ref{tab:GR_combined}. 
The $S_8$ values found in these $\Lambda$CDM-type models are compatible with the values obtained in the $F(R)$ models, shown in Table~\ref{tab:S8}. 
%, and the tensions are slightly smaller. 

%Figure~\ref{fig:fz_rg} displays the analyses of the flat-$\omega$CDM and $\Lambda$CDM-type models, i.e., 
%$\omega_0 \omega_a$CDM, and $\Omega_k$CDM, together with the 
%$\Lambda$CDM. 
%The GP reconstruction of the data agrees well with all these models within the redshift range investigated. 
%However, the $\Omega_k$CDM model consistently presents lower values 
%of $f(z)$ compared to all other models, and also with respect to the GP reconstructed function. 

%structure formation in this regime.
%
%The behavior of $\sigma_8(z)$, shown in Figure~\ref{fig:sigma8_gm}, mirrors these findings. The MG models, again, remain largely consistent with the GP reconstruction and observational data across the redshift range of the data. 
%The $\Lambda$CDM and MG models, however, predict a lower amplitude of fluctuations in the interval $0.5 \lesssim z \lesssim 1.5$, while overestimate the amplitude at $z \gtrsim 1.5$, highlighting discrepancies in the timing and strength of structure formation relative to alternative scenarios. 

\end{document}